# The behavior of grain boundaries in the Fe-based superconductors

J. H. Durrell[1], C.-B. Eom[2], A. Gurevich[3], E. E. Hellstrom[4], C. Tarantini[4], A. Yamamoto[5], D. C. Larbalestier[4],

PACS:

[1]Department of Engineering, University of Cambridge, Cambridge CB2 1PZ, United Kingdom.

[2]Department of Materials Science and Engineering, University of Wisconsin, Madison WI 53706 USA

[3]Department of Physics, Old Dominion University, Norfolk, VA 23529, USA

[4]Applied Superconductivity Center, National High Magnetic Field Laboratory, Florida State University, Tallahassee FL 32310, USA

[5]Department of Applied Chemistry, University of Tokyo, Tokyo 113-8656, Japan

## Abstract

The Fe-based superconductors (FBS) are an important new class of superconducting materials. As with any new superconductor with a high transition temperature and upper critical field, there is a need to establish what their applications potential might be. Applications require high critical current densities, so the usefulness of any new superconductor is determined both by the capability to develop strong vortex pinning and by the absence or ability to overcome any strong current-limiting mechanisms of which grain boundaries in the cuprates are a cautionary example. In this review we first consider the positive role that grain boundary properties play in the metallic, low temperature superconductors and then review the theoretical background and current experimental data relating to the properties of grain boundaries in FBS polycrystals, bi-crystal thin films, and wires. Based on this evidence, we conclude that grain boundaries in FBS are weak linked in a qualitatively similar way to grain boundaries in the cuprate superconductors, but also that the effects are a little less marked. Initial experiments with the textured substrates used for cuprate coated conductors show similar benefit for the critical current density of FBS thin films too.  We also note that the particular richness of the pairing symmetry and the multiband parent state in FBS may provide opportunities for grain boundary modification as a better understanding of their pairing state and grain boundary properties are developed.

# Glossary

AF, antiferromagnetic

A-J, Abrikosov-Josephson

BCS, Bardeen-Cooper-Schrieffer

FBO, $BaFeO_2$

FBS, Fe-based superconductors

GB, grain boundary

GL, Ginzburg-Landau

IG, intragrain

intergranular critical current density $J_{cgb}$

intragranular critical current density $J_{cg}$

Critical current density in a general sense, $J_c$

LAGB, low angle grain boundary

LSAT, $(La,Sr)(Al,Ta)O_3$

LTLSM, low temperature laser scanning microscopy

MO, magneto-optical

SF, self-field

STO, $SrTiO_3$

YBCO, $YBa_2Cu_3O_{7-x}$

# I. Introduction

The essential "big" application of superconductors is the generation of strong magnetic fields in significant volumes. As compared to Cu and Fe which are limited to about 2T, the ability to make wires that can operate at critical current densities $J_c > 10^5$ A/cm$^2$ gives a very special, "unfair" competitive advantage to superconductors that was well recognized by Kamerlingh Onnes very soon after his discovery of superconductivity. Just two years after the discovery in 1913, he described his vision for generating 10 T fields using superconducting Hg or Pb wires, a capability that he estimated would cost as much as a naval cruiser if performed with Cu cooled by liquid air, but which would require only modest resources if made from liquid helium cooled superconducting wires [1]. Sadly, almost 50 years had to pass before the crucial difference between the type I superconductors that Kamerlingh Onnes had discovered and the type II superconductors that are needed for high fields became clear. Only with intermetallic compounds such as Nb$_3$Sn and alloys like Nb47wt.%Ti would Kamerlingh Onnes's vision be accomplished [2]. A very widespread industry has developed employing these Nb-based materials, all of which is based on operation in the narrow temperature range of about 1.5-6K using helium cooling. The cost of and experimental care needed to work with liquid helium made it natural that many would dream of superconducting technology working at 77K in liquid nitrogen instead of liquid helium. Consequently immense excitement attended the discovery of superconductivity in YBa$_2$Cu$_3$O$_{7-\delta}$ with a transition temperature, $T_c$, of 92K in 1987. Why then, has it taken so long to get cuprate applications into widespread service and why is this whole issue of *Reports on Progress in Physics* devoting itself to consideration of superconductivity in the Fe-based chalcogenides and pnictides, when their $T_c$ values are well below those found in the cuprates and, so far, all below 77K? There are of course many answers, one of the most basic being that superconductivity is a property of Fe in these complex structures and magnetic Fe has long been thought to be one of the few elements intrinsically inhospitable to superconductivity. But a more practical reason is because the new class of Fe based superconductors possesses a very appealing combination of reasonable $T_c$ (up to 56K), very high upper critical field $H_{c2}$ (>100T), low $H_{c2}$ anisotropy (from about $\gamma = H_{c2}^{||} / H_{c2}^{\perp}$ = 7-9 down to as little as 1.5 or less, whereas the minimum for any cuprate is about 5), and critical current density $J_c$ well above 10$^5$ A/cm$^2$ for fields above 20T. All of this gives reasons for practical interest, well beyond their intrinsic scientific interest. But, returning to Kamerlingh Onnes's vision, we must recall that today's 10 tesla laboratory magnets of a few cm bore typically require 5-10 km of ~1 mm diameter wire. Single crystals cannot be grown in such lengths and consequently polycrystalline wires with many grain boundaries are inevitable. These grain boundaries (GBs) form a network which must be crossed by the supercurrent as it passes along the wire, making GB superconducting properties of vital importance. The unwelcome lesson of the cuprates is that current-carrying capability is seriously degraded by GBs of arbitrary misorientation, which act as weak links with intrinsically low intergranular critical current density $J_{cgb}$ much less than the intragranular critical current density $J_{cg}$. The central question addressed by this review is whether the GBs in FBS are weak links too, and, if they are, whether this is an intrinsic behavior or one induced by present-day imperfections of polycrystalline FBS material forms which will be ameliorated by clever processing.

Although virtually all semiconductor technology depends critically on the production of high purity Si single crystals in large quantities, most other useful bulk material forms are polycrystalline with collective properties determined both by grains and grain boundaries. Indeed the size of grains, their texture or local misorientation and nature (e.g. clean, wetted, or sites for segregation or second phase precipitation) are a staple component of most undergraduate materials science courses because they play a key role in determining the practical properties of many materials [3]. Sometimes it is clear that GBs are useful, for example in strengthening most metals at temperatures below about half the melting point. Sometimes, as is the case for jet engine turbine blades, GBs are deleterious. In this case, single-crystal turbine blades many cm long are made from very complex Ni-base super-alloys so that they can operate at high stress close to the melting point. Minimizing the grain-to-grain misorientation, or developing a strong texture in polycrystals, is an alternative approach when single-crystal technology is not possible. Strong polycrystalline texture is required for coated conductors of the cuprate superconductor $YBa_2Cu_3O_{7-x}$ (YBCO) or for grain-oriented [4] Fe-3%Si steels, where low hysteresis loss sheets for transformer cores are stamped so as to allow orientation of the flux density along the easy axis of magnetization [5]. The greatest challenge to the application of cuprate superconductors has been the need to develop texturing processes that approximate single crystals by the mile. The question of whether the same technology needs to be applied to FBS is a natural and vital question that forms a concluding emphasis of this paper.

Under any particular set of conditions, the primary figure of merit for applications of a superconductor is its critical current density $J_c$, normally defined by the measured critical current $I_c$ divided by the cross-section $A_{sc}$ of superconductor. Here $I_c$ is simply the current that can be supported before resistance is observed. $I_c$ is finite because the transverse Lorentz force density ($\mathbf{F_L} = J_c \times \mathbf{B}$) that would dissipatively move the flux lines which penetrate Type-II superconductors is opposed by a pinning force $F_p$ exerted by the interaction between the vortices and microstructural defects [6]. This pinning force $F_p$ is primarily dependent on interactions between either the normal-state vortex cores or their screening currents and defects where superconductivity is suppressed. As grain boundaries are lattice defects, they too can perturb the local superconducting properties, affecting both the primary superconducting properties such as $T_c$ and the thermodynamic critical field $H_c$, but also the local superconducting properties that control both the local pinning force, sometimes called the elementary pinning force $f_p$, and its global summation $F_{pmax}$ over all defects and vortices.

A key point that provides background to the discussion of this paper is that, although there are thousands of superconductors, so far there have been just 6 materials that have advanced into serious production as conductors from which superconducting magnets can be wound. Three of these are electron-phonon, high carrier-density, good-metal superconductors, Nb-Ti [7], $Nb_3Sn$ [8] and $MgB_2$ [9]. In these materials GBs do have slightly depressed superconducting properties, but the depression is small enough as to allow significant pinning of vortices to GBs without inhibiting the flow of current across the GB. In fact GBs are the dominant pinning center in A15-phase ($Nb_3Sn$) [10], Chevrel-phase ($SnMo_6S_8$) [11] and $MgB_2$ [12]. For the three cuprate compounds that have made it into production as conductors for magnets, $Bi_2Sr_2CaCu_2O_{8-x}$, $(Bi,Pb)_2Sr_2Ca_2Cu_3O_{10-x}$ [13] and $YBa_2Cu_3O_{7-x}$, [14] grain boundaries are always obstructive at large misorientations, making the effective cross-section for percolative current flow $\ll A_{sc}$. Studies of single GBs using simple, planar GBs produce a rather general result that the GB critical current density $J_{cgb}(\theta)$ falls off exponentially when the GB misorientation $\theta$ exceeds a critical angle $\theta_c$, where $\theta_c$ may be as little as 3-5° for planar GBs [15] [16]. But $J_{cgb}$ across GBs in "real" cuprate conductors are often better than the planar "science" GBs of epitaxially grown bicrystals [17] for reasons that may also be valid for FBS polycrystals. Since a central conundrum of modern superconductor science is whether the drive to higher $T_c$, more complex and less metallic structures is necessarily one that also makes the superconducting state more sensitive to local variations of structure and composition that bring with them obstructive GBs, we take the task of understanding GBs in cuprates and now FBS as being both a central scientific *and* a central technical question for modern superconducting materials research.

Our paper is organized along the following lines. In section II we first discuss some basic issues about grain boundaries in the more metallic, lower $T_c$, and the newer, higher $T_c$, less metallic superconductors. We describe the central reasons why GBs in cuprates exhibit strongly depressed superconductivity and then draw on common features of cuprates and the FBS to suggest why similar behavior is to be expected there too. We describe some of the very attractive intragrain properties of the FBS and then address ways to extract some knowledge of the intra- and inter-grain properties from studies of polycrystals. Polycrystalline studies remain important because high quality thin films have not yet been made for many of the FBS materials and single GB studies have been made on even fewer. In section III we address the experimental results obtained on thin film bicrystals, polycrystalline bulks and polycrystalline wires, while in section IV we provide a discussion and summary of our findings.

## II Basic Issues

To allow the reader to appreciate the importance of grain boundaries in FBS we first provide some general background on grain boundaries, contrasting the behavior of GBs in low temperature, high carrier density, metallic superconductors like $Nb_3Sn$ and the Chevrel phases with that found in the cuprates, especially $YBa_2Cu_3O_{7-x}$. Because measurements of single GB properties have been made only for a small fraction of the FBS compounds, we also describe experimental techniques and some of the complexities involved in deriving grain boundary properties from measurements of polycrystals.

### II.1 Grain Boundaries in Superconductors

Grain boundaries naturally arise as a consequence of there being many nuclei that can initiate transformation of the precursor phase into the desired phase. GBs occur at the impinging interfaces between growing grains and they thus form a 3D network across which any long range transport has to occur. The structurally disordered region at the GB is narrow, typically 1 nm thick. It will be seen later that the superconducting properties of GBs are determined by the relative dimensions of several nanometer-scale parameters. First is the length scale of the structural disorder, non-stoichiometry and lattice strains, second is the Thomas-Fermi screening length, and third is the superconducting coherence length $\xi$, which quantifies the size of the Cooper pair. To describe GBs geometrically requires defining the plane of the GB and the misorientation angle $\theta$ (see [18] for a full discussion of how to specify uniquely the crystallographic relationships between neighboring grains). In general, metallic superconductors made in bulk have little or no texture and strongly curved GBs. By contrast, attempts to understand the properties of GBs in the cuprates have emphasized texture development so as to minimize the misorientation $\theta$ and attempts to favor planar GBs of high symmetry whose properties are much more uniform than is the case with curved, non-planar GBs.

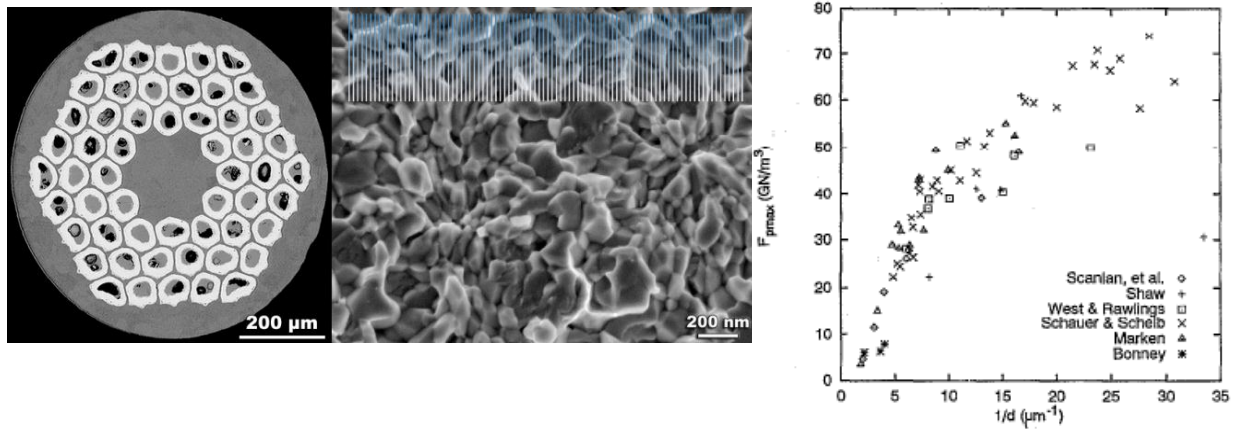

Figure 1. At left is a modern conductor containing many thousand $Nb_3Sn$ filaments and at center the $Nb_3Sn$ grain structure (images courtesy Peter Lee at the, NHMFL). The curved, non-planar nature of the GBs is evident. Although $F_{pmax}$ occurs at about 5T and 4.2K, the density of GBs is several times less than the vortex density at 5 T, which is indicated by the light grey lines superimposed on the micrograph. At right is shown the relationship between grain size and $F_{pmax}$(4.2K) in $Nb_3Sn$ conductors of various types and in $SnMo_6S_8$ (Reprinted with permission from [11]. Copyright 1995, American Institute of Physics)..

The most general result for metallic superconductors is that the higher the density of GBs, the higher is the critical current density. Figure 1 illustrates the cross-section of a state-of-the-art, multifilamentary $Nb_3Sn$ conductor about 1 mm in diameter that carries about 1000 A, and a fractograph obtained by breaking the filaments and observing the largely intergranular fracture surface on which the grain size is seen to be 100-200 nm in diameter. The maximum pinning force density $F_{pmax}$ of $Nb_3Sn$ can be changed by varying the temperature at which the $Nb_3Sn$ is formed. It is universally [19] [20] [21] found that higher temperature formation means larger grains and a lower value of $J_c$ and $F_{pmax}$. The inverse relation with grain size of Figure 1 in fact means that $J_c$ is proportional to the area of GB per unit volume. This relationship holds for single-phase Nb-Ti [22], for the Chevrel phase superconductor $SnMo_6S_8$ [11], and for $MgB_2$ [23] [12]. Two conclusions may be drawn. One is that GBs in metallic superconductors cannot be intrinsic barriers to supercurrent flow, even if, as in $MgB_2$, there is abundant evidence that insulating MgO at GBs can partially obstruct the GBs [24]. The second is that GBs must actually enhance the vortex pinning capability of the superconductor by having properties that are at least partially degraded with respect to the grain interiors so that there is a binding energy of vortices to the GB.

The scientific principles behind GB vortex pinning were laid out in various papers in the 1980s [25] [26] [27]. The authors all use first order perturbations of the Ginzburg-Landau (GL) free energy in the vicinity of a GB to calculate the pinning interaction between a vortex and a GB, the mechanism being the change in electron mean free path $\ell$ due to elastic scattering at the GB. This scattering changes the GL parameter $\kappa = \lambda/\xi$, where $\lambda$ is the superconducting penetration depth and $\xi$ is the superconducting coherence length, through its effect on the impurity scattering parameter $\alpha = 0.882$ $\xi_0/\ell$, where $\xi_0$ is the BCS coherence length in the clean limit ($\xi_0 << \ell$). Welch [28] then amplified this treatment to take account of enhanced vortex pinning by thickening the GB as may occur in $Nb_3Sn$ where Cu can segregate to the GBs [29]. These studies remain valid today: the technology of developing high $J_c$ in Nb-Ti and $Nb_3Sn$ depends on the fact that their GBs are beneficial. At minimum there is no evidence that their presence harms supercurrent transport and, especially for $Nb_3Sn$, it is quite clear that it is essential to maintain the finest grain size possible so that vortex pinning at the GBs achieves its maximum effect. Neither of these statements is true for the cuprates, as became very clear when polycrystals of the cuprates were examined early in 1987 and found to have very low $J_c$ [30].

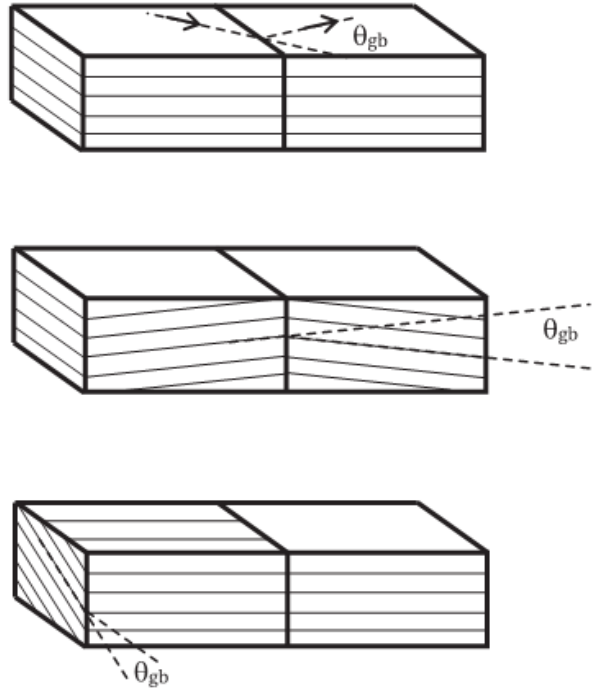

**Figure 2.** Types of GB misorientation θ found in common bicrystal substrates used for thin film growth. From top to bottom are shown an [001] tilt boundary, a [100] tilt boundary and a [100] twist boundary. In this special geometry, epitaxial growth of the superconducting overlayer tends to produce a planar GB that replicates the substrate geometry. GBs formed in bulk samples seldom have such simple, high symmetry crystallographic relationships, nor are their GBs confined to a single plane, as shown here. In polycrystals, the curved planes shown in Figure 1b are more common.

Before addressing the superconducting properties of GBs in the cuprates and FBS, let us return to the general properties of GBs as they manifest themselves in layered materials like the cuprates. Anisotropic growth is common in layered, quasi-2D phase like the FBS and cuprates, often producing weak growth textures with slightly preferred values of θ that tend to an average lower than random. In cuprates it was very soon seen that texture was helpful in raising the $J_c$ evaluated on scales much larger than the grain size. The natural next step was to look for special grain-to-grain misorientations with low angle and high symmetry.

The first convincing proof that high $J_c$ was possible in cuprates came when epitaxial thin film growth methods were developed [31] using perovskites like SrTiO$_3$ as the epitaxial substrate on which to grow YBa$_2$Cu$_3$O$_{7-x}$. This very quickly led to the production of bicrystal perovskite substrates by sintering two such single crystals together [32] and the growth of thin film bicrystals on such SrTiO$_3$ bicrystal substrates. The crystallographic relationships that underpin most scientific studies of single GBs in both cuprates and FBS are shown in Figure 2. When the plane of the GB contains the rotation axis, the misorientations are tilt misorientations but when the rotation axis lies perpendicular to the GB plane the GB is described as having a twist rotation. Thus Figure 2 contains from top to bottom an [001] tilt, a [100] tilt and a [100] twist GB. In the top and bottom cases, the [001] direction (or the *c*-axis) is perpendicular to the top surface of the film. Although these planar, high-symmetry GBs are unusual in any sample in which independent nucleation occurs randomly (i.e. in most bulks and *ex situ* films), they are favored by *in situ* growth seeded by the substrate, where they have been hugely important for providing a basic understanding of what causes the GB to degrade with respect to the grains. Especially for low angle GBs (LAGBs) where the GB spontaneously subdivides into a chain of dislocations and channels of only slightly distorted crystal, the mix of properties provided by channel and dislocation is both of great scientific and great technological interest. LAGBs with θ <10° form a particularly simple system because the separation *d* of the dislocations is given by the Frank formula $d = b/2\sin(\theta/2)$, where b is the Burger's vector of the dislocation. As envisaged by Chisholm and Pennycook [33], the onset of weak link behavior occurs at a critical angle $\theta_c$ ~5-7° when the channel closes, but actually most studies find $\theta_c$ to be about half this, a result that emphasizes not only the need for a full understanding of local GB structure but also the importance of the interaction of vortices at the GB with the pinned vortices in the grains [34].

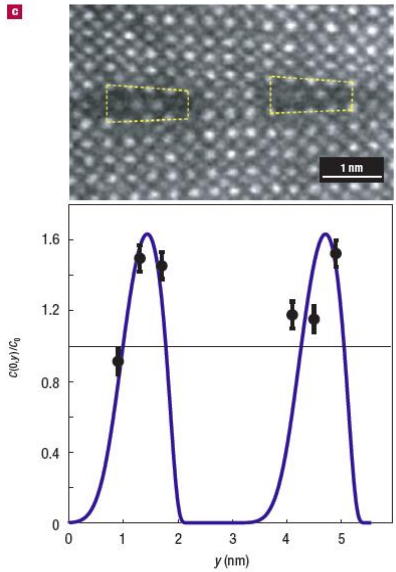

Figure 3. The upper image shows a Z-contrast view of dislocation cores at a 7° [001] tilt GB formed in a Ca-doped YBCO thin film grown on a 7° [001] SrTiO$_3$ bicrystal substrate. The bottom plot compares the measured Ca segregation enhancements that rise more than 50% above that predicted by the strain and charge segregation model discussed in section II.2 [35]. The segregation plays a beneficial role in enhancing the superconducting transparency of the GB and raising its intergranular $J_c$.

In low and intermediate angle GBs, the vortices easily become hybrids, neither Abrikosov vortices nor the Josephson vortices found at a well-developed weak link [34],[36]. These hybrid Abrikosov-Josephson (A-J) vortices are about the size of the conventional Abrikosov vortex but, instead of normal cores, they have Josephson cores elongated along the GB. As a result, such vortices are less strongly pinned than the intragranular Abrikosov vortices. Epitaxial films examined with the external field lying parallel to the GB plane very often show GB flux flow occurring before it occurs in the grains, i.e. $J_{cg} > J_{cgb}$. Thus low angle YBCO grain boundaries can also be usefully viewed as planes of weak pinning [37] [38], since it means that one route to higher $J_{cgb}$ lies through better pinning of these GB A-J vortices. Routes to achieving this include pinning by the dislocations that make up the LAGB [39], grain vortex-GB vortex interactions, especially at high fields [40], meandered grain boundaries that ensure that any one vortex has only a small length crossing the GB [41] [42] [43], and changing the angle between the applied field and the GB plane so as to prevent vortices lining up with the GB (the weak pinning plane) [43]. These techniques are distinct and additional to those that seek to enhance the grain boundary order parameter by overdoping the grain boundary, for example with calcium [44] [35]. An important point to take away from this discussion is that planar GBs provide the most easily interpreted way to study the depression of the GB properties – but actually, real, non-planar or meandered GBs of similar misorientation $\theta$ can have markedly better properties because only short segments of any one vortex lie in the GB.

Some special types of boundary can enhance the properties too as Palau *et al.* [45] have shown for a twin plane. At a twin plane in YBCO, the slightly different *a* and *b* axis lengths swap and there is a temperature-driven cross over between vortex pinning and vortex channeling because the superconducting coherence length ξ varies with temperature. At high temperatures ξ is large enough to span the region of structural disorder at the GB and the boundary can act as a pinning center, while at lower temperatures the shorter ξ means that the GB acquires depressed superconducting properties and then allows vortex channeling along GB.

Two additional points may be made before closing this section. It is often said that the root of the weak link problem in cuprates derives from their short coherence lengths. In fact, the coherence length of Nb$_3$Sn in the zero temperature limit is 3 nm, which is the same as in YBa$_2$Cu$_3$O$_{7-x}$ at 77K for H parallel to the *c*-axis. The second point is that segregation may occur at GBs, which may affect the GB properties. As envisaged by studies of Cu segregation to $Nb_3Sn$ GBs, substantial Cu enrichment can occur over depths of up to 10 nm at GBs without turning the GBs into weak links [29] [28]. As noted in the next section in our discussion of the theoretical underpinnings of the problem for FBS compounds, the key issues are the proximity of the superconducting state to the antiferromagnetic state in cuprates and FBS and their low carrier density, because this determines how well local electrostatic charges of non-superconducting GB regions, for example dislocation cores, can be screened. As Figure 3 shows, simple Z-contrast images of the GB may not be enough to understand what is occurring at the GB and attention may also need to be paid to the strong segregations of the sort that have been shown at cuprate GBs. An important case is that of partial Ca-substitution for Y in YBCO, where additional carriers are added, allowing the overdoped state to be accessed [44] [35] [46]. This is very valuable for enhancing $J_{cgb}$ and such overdoping may open possibilities for ameliorating the properties of FBS GBs too. This complex interaction has been the subject of much investigation [15-17,25-27,33,34,47,48] and its theoretical basis is discussed in the next section.

## II.2 Features of Grain Boundary Properties in FBS and Cuprates.

Both cuprates and FBS are layered materials in which superconductivity occurs primarily on specific atomic planes (the Cu-O planes in the cuprates and the Fe-As or Fe planes in FBS). These are both layered structures with considerable charge transfer between the layers. Superconductivity in the cuprates occurs on doping a Mott antiferromagnetic (AF) insulator [49] while FBS become superconducting by doping a parent AF semi-metal [50] [51] [52], which means that superconductivity competes with AF states in both cases. The similarities of cuprates and FBS become even more apparent when comparing their phase diagrams, in which AF and superconducting domes appear in a similar manner on increasing the hole or electron concentration, x, as shown in Figure 4.

Quite generally, these diagrams suggest that if the GB structure causes local reduction of x due to local non-stoichiometry, strain or charge effects (see below), superconductivity becomes suppressed and the parent AF phase may precipitate at the GB or other crystal defects, giving rise to local weak link behavior. In addition to the unconventional and not yet fully understood microscopic mechanisms of superconductivity, the cuprates and FBS have other remarkable similarities, which may cause problems for applications:

1. They both have high normal state resistivities, low carrier densities, and low Fermi energies compared to conventional superconductors. As a result, both FBS and the cuprates have very short coherence lengths, $\xi_0$ ~1-2 nm inferred from their extremely high upper critical fields $H_{c2} = \phi_0/2\pi\xi_0^2$ [53], but comparatively large Thomas-Fermi screening lengths $l_{TF} \sim \xi_0$. Here $\phi_0$ is the flux quantum, 2.07 x $10^{15}$ Wb.

2. The superconducting state is in close proximity to competing AF states.

3. The unconventional (i.e. not single-band s-wave) symmetry of the Cooper pairs, d-wave in the cuprates and multiband s-wave with a possible sign change of the superconducting gap on disconnected pieces of the Fermi surface in FBS (the so-called $s^{\pm}$ pairing), may impose barriers to the continuity of the superconducting phase at even small structural perturbations [54].

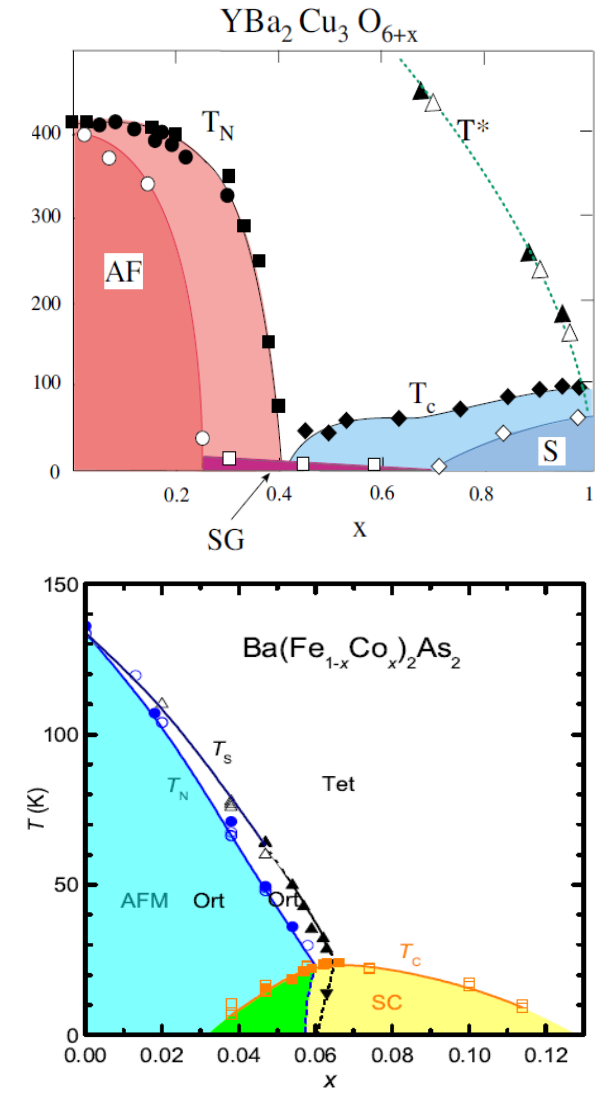

Figure 4. Similarities of the phase diagrams of cuprates and Fe-based superconductors. The upper panel shows the phase boundaries between the superconducting (S) and anti-ferromagnetic (AF) phases for the optimally doped and 4%-Zn substituted YBCO as functions of the hole concentration x. Zn substitution suppresses $T_c$ and expands the AF state (Reprinted from [55], Copyright 2000 by Physical Society of Japan). Lower panel shows the AF and superconducting (SC) states in Co-doped 122 FBS (Reprinted figure with permission from [56]. Copyright 2010 by the American Physical Society). The structural change from orthorhombic (Ort) to tetragonal (Tet) is also indicated.

4. Large ratios of the electron masses along the c-axis and the ab-plane $\gamma^2 = m_c/m_{ab}$, with $\gamma$ ranging from ~ 1 to ~ 50 for FBS and from ~ 20 to > $10^4$-$10^5$ for the cuprates, make transport through polycrystals difficult.

5. Their complex chemical compositions and directional bonding can cause great sensitivity of the local superconducting properties to local structural imperfections.

Points 1 and 4 enhance thermal fluctuations which significantly reduce the useful T-H domain in which superconductors can carry high current densities, even in the presence of strong pinning [57]. Points 1-5 can result in weak linked grain boundaries which impede current flow because the critical current density through a grain boundary $J_{cgb}(\theta) = J_0 \exp(-\theta/\theta_c)$ drops exponentially as the misorientation angle $\theta$ increases [15]. In the cuprates $\theta_c$ ~ 3-5° so the spread of misorientation angles ~40° in polycrystals can reduce $J_{cgb}$ by 2-3 orders of magnitude. Recent experiments revealed similar weak linked grain boundaries in $Ba(Fe_{1-x}Co_x)_2As_2$ bicrystals with $\theta_c$ ~ 10° [58] [59] and polycrystalline $LaFeAsO_{1-x}F_x$ [60]. Yet it appears that neither a comparatively high $T_c$ nor multiband superconductivity is by itself sufficient to force GB weak link behavior. For example, grain boundaries in the two-band $MgB_2$, whose $T_c$ = 40 K is higher than the $T_c$ of 122, 111 and 11 FBS are not weak links [61]. However, $MgB_2$ is a conventional electron-phonon superconductor with $\xi_0 \gg l_{TF}$ and none of the features outlined in points 1-5.

The d-wave symmetry of the order parameter in the cuprates can result in the formation of low-energy Andreev bound states at a continuous interface between misoriented crystallites. The d-wave Andreev states form at certain orientations, weakening the superconductivity at those interfaces [62,63], which primarily occurs on high-angle GBs. However, low-angle GBs are not continuous interfaces since they are formed by chains of dislocations spaced by the distances d = b/2sin(θ/2) which can be greater than the coherence length $\xi$ for small $\theta$. In this case the Andreev states do not form in the undisturbed current channels between the GB dislocations (although they might form at the dislocation cores), so the role of Andreev states on the current transport of low-angle GBs is diminished. The multiband FBS are likely not d-wave and the physics of the Andreev bound states across interfaces in multiband $s^{\pm}$ superconductors and their role on current transport across GBs in FBS have not yet been addressed.

The first SmFeAsO$_{1-x}$F$_x$ [64] [65], (Sr$_{0.6}$K$_{0.4}$)Fe$_2$As$_2$ [66] and Fe(Se,Te) wires made by the *ex situ* powder-in-tube method [67] exhibited poor grain connectivity and had rather low J$_c$(5 K, 1 T) ~ 10 – 10$^3$ A cm$^{-2}$. Grain boundaries in these wires are likely coated by non-superconducting second phases, as has been revealed by electron microscopy of FBS polycrystals [68] [69] [70] [71]. Assuming that further technological refinements will eliminate such impurity-phase-induced granularity, porosity and other extrinsic factors, the fundamental question remains: Are clean grain boundaries in FBS intrinsic weak links as they apparently are in the cuprates? So far, clean current-limiting grain boundaries have only been observed in Ba(Fe$_{1-x}$Co$_x$)$_2$As$_2$ bicrystals [58] [59], so it may be premature to conclude that intrinsic weak links are characteristic of all FBS.

From a fundamental point of view, it appears that the GB problem in both cuprates and FBS may be related to the mechanisms which provide high values of $T_c$ and $H_{c2}$. Indeed, low carrier densities, short coherence lengths, unconventional pairing symmetry, large screening lengths and competing AF states are characteristic of both cuprates and FBS and surely do contribute to the suppression of superconductivity at grain boundaries [15] [72] [48] [35] through various intrinsic mechanisms. Among those are: 1. Strains near GB dislocation cores which can transform them into current-blocking regions [72]; 2. Space charges caused by charged dislocation cores or nonlinear elastic deformations at the GB [72] [48]; 3. Impurity segregations (Cottrell atmospheres) induced by the strain and electric field of the GB [35]; 4. Competing current-blocking AF phase (see Figure 4) can be induced at the GB by strain, charge and impurity effects.

The effect of GB strains on superconducting properties is quantified by the values of $\partial T_c/\partial \varepsilon$ where $\varepsilon_{ij}(\mathbf{r})$ is the strain tensor produced by GB dislocations or facets. Because of the comparatively large $\partial T_c/\partial \varepsilon \approx$ 200-300K even in the optimally-doped cuprates, the strains $\varepsilon_{ij}(r) \approx b/2\pi r$ around dislocation cores can locally suppress superconductivity at distances r of the order of a few Burgers vectors b [72]. In FBS not only is the strain dependence of T$_c$ significant, but T$_c$ degrades as the bond angle $\alpha(\mathbf{r})$ in the Fe-As tetrahedrons deviates from the ideal angle of 109.5$^o$ [52]. Thus, tilt and shear distortions of $\alpha(\mathbf{r})$ at the GB may depress superconductivity and block current even at LAGBs. Another distinctive feature of FBS is a giant magneto-elastic coupling [52] which may result in the nucleation of magnetic structures by GB dislocations or facet strains. Spin-flip scattering at such strain-induced magnetic structures could also contribute to the deterioration of current transport across FBS GBs.

Charging effects at the GB result from either atomic charges at dislocation cores (common in many complex compounds [73] [74] and also in the cuprates [48] [35] [75]) or anharmonic expansion of the lattice due to alternating strains around dislocations [72]. The resulting shift of the mean electric potential $\varphi_0$ on the GB is defined by a solution of the Thomas-Fermi equation, which describes the screened electric potential $\varphi(x) = (2\pi q l_{TF}/sd\kappa)\exp(-|x|/l_{TF})$ across the GB, where q is the mean dislocation charge per *ab* plane, s is the spacing between the *ab* planes, d = b/2sin($\theta$/2) is the spacing between the dislocations along the GB [73], and $\kappa$ is the dielectric constant due to filled electronic bands. The screening space charge at a GB results in electron band bending, shifting the chemical potential $\mu$ of the GB by the value $e\varphi_0 = (4\pi q e l_{TF}/sb\kappa)\sin(\theta/2)$ increasing as the misorientation angle increases (here $e$ is the electron charge). For $s \approx b \approx 0.5$ nm [50] [51] [52], $l_{TF}$ = 0.2 nm, q = e, $\kappa$ = 10, and $\theta$ = 20°, we obtain $e\varphi_0 \approx 250$ meV, which is much smaller than the Fermi energy $E_F$ of metallic superconductors like Nb, but *larger* than $E_F$ of the semi-metal FBS, particularly the 122 and Fe(Se,Te) chalcogenides for which ARPES studies show rather small Fermi energies $E_F$ = 20-50 meV [52] [76]. These estimates suggest that charged dislocation cores in FBS can indeed turn the GB antiferromagnetic or even insulating. The actual superconductivity depression on the GB can be evaluated using the Ginzburg-Landau (GL) equation $\xi_0^2 \psi'' + [\tau - \Gamma_0 \delta(x)]\psi - \psi^3 = 0$ where $\tau = (T_c - T)/T_c$ and the parameter $\Gamma_0 = e\varphi_0 l_{TF} \partial \ln T_c/\partial \mu$ quantifies the reduction of $T_c$ due to the charge-induced shift of the chemical potential $\mu$ at GB. The solution of the GL equation [72] yields the following ratio of the superconducting gap $\Delta_0$ to the bulk gap $\Delta$:

$$\frac{\Delta_0}{\Delta} = \frac{1}{\sqrt{1+\Gamma^2}+\Gamma}, \qquad \Gamma = \frac{2\pi q e l_{TF}^2}{\tau s b \xi_0 T_c \sqrt{2\tau}} \left[\frac{\partial T_c}{\partial \mu}\right] \sin\frac{\theta}{2} \qquad (1)$$

Reduction of the gap $\Delta_0$ at GBs becomes particularly pronounced at $\Gamma > 1$, because $\Gamma$ is increased for short coherence lengths $\xi_0 \approx \hbar v_F/2\pi k_B T_c$, for long screening lengths $l_{TF}$ and for large values of $\partial T_c/\partial \mu \sim T_c/E_F \sim 0.4$-$0.7$ K/meV, all of which seem to be characteristic of FBS. These features of the semi-metallic FBS naturally result from their low carrier densities n, small Fermi energies and comparatively high $T_c$. For instance, $l_{TF} = [\kappa/4\pi e^2 N(E_F)]^{1/2}$ for a weakly isotropic semi-metal can be estimated using the 3D density of states $N(E_F) = m^2 \gamma^{1/2} v_F/\pi^2 \hbar^3$ for which $l_{TF} \propto n^{-1/6}$ increases as n decreases. However, in the 2D layer limit, where $\gamma \gg 1$, the screening length $l_{TF} = (sr_B\kappa)^{1/2}/2$ becomes independent of n where $r_B = \hbar^2/me^2$ is the Bohr radius.

The generic phase diagrams of cuprates and FBS shown in Figure 4 suggests that competition of superconductivity and antiferromagnetism always has the potential to nucleate the parent AF phase around dislocation cores by small charge/strain-driven shifts of the chemical potential toward the AF state. As discussed above, the very small Fermi energies of FBS and the resulting strong band bending at GBs can thus make polycrystalline FBS prone to this intrinsic weak link behavior. Another interesting feature of GBs in FBS may result from the overlapping AF and superconducting domes in Figure 4, suggesting that the AF phase in 122-FBS (unlike the case of YBCO) may be present at GBs even in the superconducting state. These peculiarities of FBS may result in a rich variety of strain, charge and spin structures coexisting with the superconducting state, which can affect GB supercurrent transport.

The GB supercurrent can also be affected by the sign of the dislocation core charge *q*. For instance, if *q* has the same sign as the charge e of the dominant carriers, the GB may cause local depletion of the carrier density, suppressing superconductivity and nucleating very local AF phase structures. This situation is apparently characteristic of GBs in YBCO where a negative potential of − 2.4 V characteristic of a locally depleted hole content was inferred from electron holography measurements on a 4° tilt GB [75]. In the case of opposite signs of *q* and *e*, the GB may enter an overdoped state (if the grains are optimally doped), which can also depress superconductivity according to the dome-like phase diagram in Figure 4, but without the generation of local AF or dielectric phases. However, FBS are multiband superconductors with several electron and hole pockets in the Fermi surface, so a charged GB may deplete the carrier density in one band and increase it in the other. The opposite response of hole and electron pockets to charged GBs may make understanding the GB transport in FBS even more complicated as compared to the single-band cuprates.

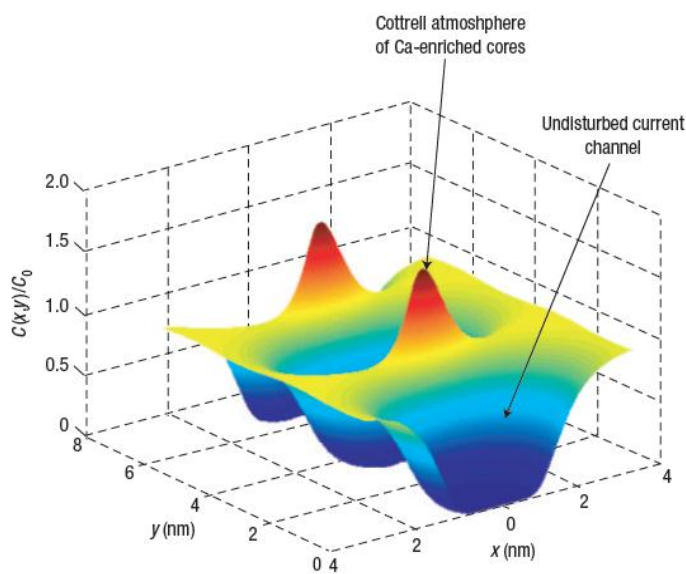

Figure 5 Predicted segregation of impurity atoms c(x,y) due to the strain and electric fields of charged GB edge dislocations. The calculations describe the equilibrium distribution of $Ca^{2+}$ solute ions around a 7°[001] tilt GB of a Ca-doped $YBa_2Cu_3O_7$ bicrystal [35]. See also the experimental measurements in Fig. 3 which provided the first experimental validation of this model. The films also show an enhancement of the intergranular $J_c$.

The sensitivity of phase diagrams of cuprates and FBS shown in Figure 4 may be particularly characteristic of chalcogenide FBS because of their small Fermi energies $E_F \sim$ 20-50 mV [52] [76]. Generally speaking, band-bending effects become more pronounced in underdoped states where carrier density is low and screening of local charges produced by GB structural disorder becomes less effective. In the cuprates the weak link behavior of underdoped GBs can be partially improved by Ca overdoping (although at the expense of reduced $T_c$ in the grains) [15]. However, segregation of impurities in the strain and electric fields of charged GB dislocations occurs in a highly inhomogeneous manner: impurities with atomic volume $V_0$ smaller than the volume V of host atoms tend to cluster in compressed regions of the lattice [35]. Such Cottrell atmospheres of impurities [73] [74] form in the lattice pressure field $p(x,y)=p_0\sin(2\pi y/d)/[\cosh(2\pi x/d)-\cos(2\pi y/d)]$ composed of dipolar strain fields of GB dislocations, p(x,y) oscillating along GB (y-axis) and decaying over a length $d/2\pi$ across the GB along the x-axis in Fig. 5. Here $p_0 = \mu\sin(\theta/2)/(1-\nu)$ where $\mu$ is the elastic shear modulus and $\nu$ is the Poisson ratio [73]. Moreover, a charged GB can impose a Coulomb barrier for segregation of impurities having charge of the same sign as that of the dislocation cores, while encouraging impurity segregation of opposite charge. An example of the experimentally observed and calculated equilibrium distribution of impurities around a Ca-doped YBCO low-angle GB [74] is shown in Figure 3, while Figure 5 shows the full 3D calculation. It is clear that such strain and charge driven segregations can be large and fully capable of supporting the rich variety of electronic states discussed above. Fortunately, such charge segregation can enhance the properties of cuprate GBs [35], giving some hope that similar effects may be possible for FBS GBs too.

Given that superconductivity in the FBS originates from the d-orbitals of the quintessential Fe ions, grain boundaries in addition to the charge/strain structures described above may also exhibit interesting spin structures, which might manifest themselves by precipitation of magnetic oxide phases or segregation of Fe vacancies with magnetic moments at the GB [37]. However, the strain/charge mechanisms described above can also give rise to intrinsic AF structures at GBs. In turn such structures can add more complexity to the impurity segregation around GBs, particularly for paramagnetic impurities which can also interact with the exchange fields of GB magnetic structures. In short, the variety of behavior that is possible at FBS GBs could be extraordinarily rich. Sadly, few of these structural changes are likely to enhance the superconductivity of the GB.

## II.3 Intragrain Properties

Before embarking on a study of grain boundary properties in the FBS, it is first of all useful to understand their intra-grain properties. The physical properties of Ba-122 epitaxial thin films deposited by pulsed laser deposition using template engineering are a convenient system to take as an example due to the availability of high quality thin films made by this route.

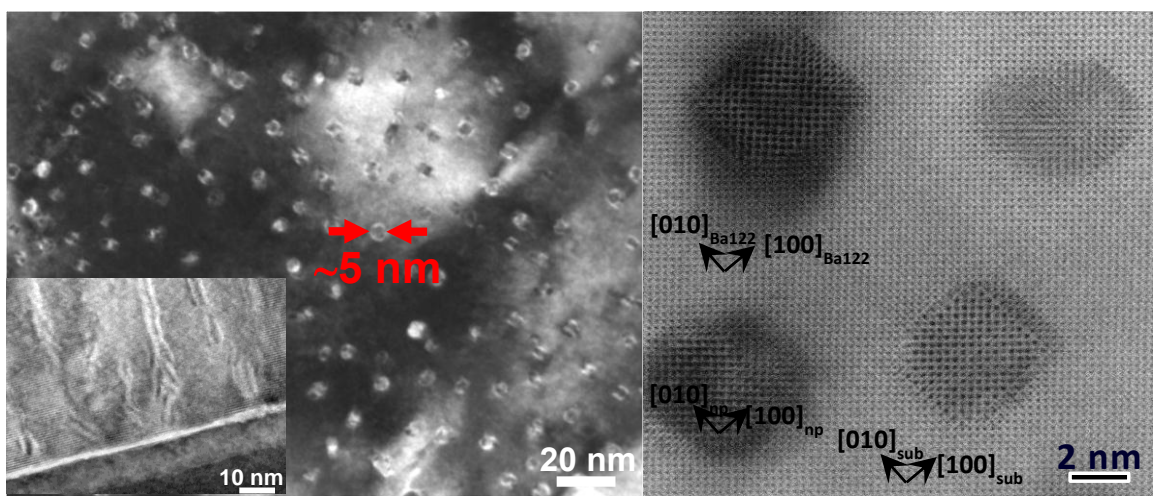

**Figure 6.** At left, planar view of the dense $BaFeO_2$ nanorods lying in the matrix of the Co-doped Ba122 thin film. Inset at lower left shows a longitudinal section of nanorods and matrix in which the vertical continuity of the nanorods is clear [77]. An atomic resolution plan-view image showing the excellent atomic plane continuity between nanorod and matrix and the very small distortions is at right [78].

Due to the volatility of several elements that play important roles in FBS materials (e.g. As, F, K, O etc.), $Ba(Fe_{1-x}Co_x)_2As_2$ has proved to be the easiest system from which to make good quality thin films, primarily because the doping element Co is not volatile. However, compatibility with the usual oxide substrates turned out to be an important consideration. Without trying to be exhaustive, we here describe the properties of 350nm Co-doped Ba-122 films grown on (001)-oriented $(La,Sr)(Al,Ta)O_3$ (LSAT) substrates using an intermediate template of 50-150 unit cells (u.c.) of $SrTiO_3$ (STO), since the template induces a full in-plane orientation without producing weak linked GBs [58] [79]. Such templates allow films of very high epitaxial quality, but microstructural characterization by transmission electron microscopy (TEM) revealed the formation of randomly distributed columnar defects that yield very strong flux pinning. Their composition was identified by electron energy-loss spectroscopy as $BaFeO_2$ (BFO) [78], suggesting a beneficial oxygen contamination of the growing film either from the PLD chamber vacuum or from residue in the target. In Figure 6 (left panel) the planar view of a 122 film grown on 100 u.c. STO/LSAT shows a high density of defects with average diameter of 5 nm. The inset shows that these columnar defects are vertically aligned and grow from the template with only a slight meandering. A mean separation of 16-17 nm was estimated from the TEM planar view, corresponding to a matching field of about 7-8 T and a volume

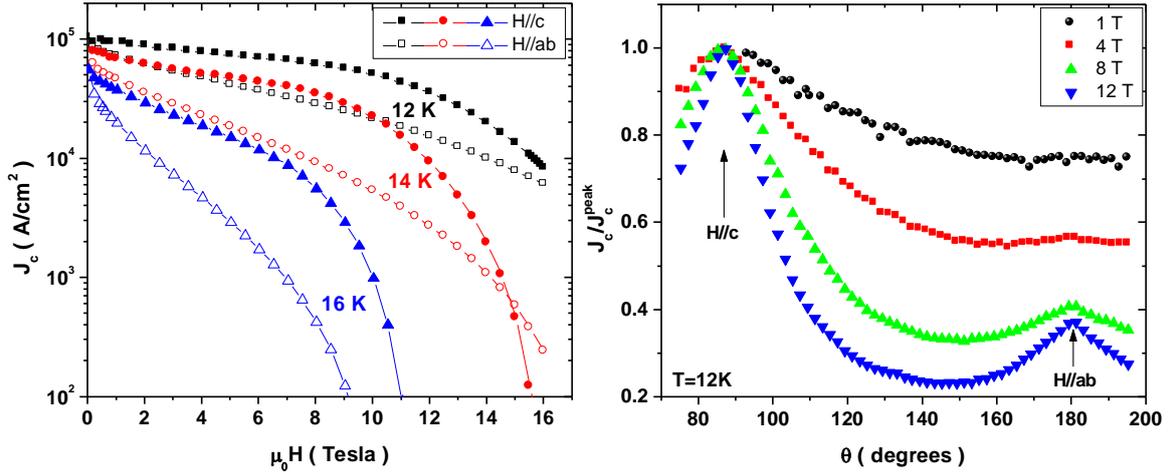

Figure 7 Critical current density $J_c$ versus applied field (left panel) and angle between H and the film plane (right panel) in an epitaxial, single-crystal Co-doped Ba-122 thin film [77]. The strong pinning produced by the BaFeO$_3$ nanorods of Fig. 6b inverts the expectation of $J_c$, raising that for H//*c* above that for H //*ab*.

fraction of about 8% [77]. This value is twice what YBCO can accept without a significant $T_c$ suppression, but is still smaller than the critical concentration of ~10% above which current blocking by pins starts decreasing the pinning force maximum $F_{pmax}$ and $J_c$ [80]. This still leaves scope for Ba122 thin films to improve. Moreover, in contrast to what occurs in the case of YBCO with BaZrO$_3$ nanoparticles, the inclusion of the BFO secondary phase nanorods in Ba-122 does not induce much stress or buckling in the superconducting matrix, as shown in the right panel of Figure 6. The important, and indeed very positive, conclusion to be drawn from this is that the Ba(Fe$_{1-x}$Co$_x$)$_2$As$_2$ structure is very accepting of second phase nanostructures, which can be very beneficial to the transport superconducting properties

The field and angular dependencies of the critical current density $J_c$ for the same film are shown in Figure 7. In the left panel the data reveal that $J_c$ is significantly larger when the field is along the c-axis instead of along the *ab*-plane up to high fields. This behavior is opposite to that expected from the mass anisotropy and indicates the presence of strong correlated pinning along the c-axis clearly related to the BFO nanorods observed by TEM. Only when the irreversibility field, $H_{Irr}$, is approached does a crossover between $J_c$(H//*ab*) and $J_c$(H//*c*) occur, restoring the expected anisotropy. $H_{Irr}$ at 12 K was estimated by Kramer plot extrapolation to be 20 and 24 T for H//*c* and H//*ab*, respectively, a very impressive number considering that $T_c$ is only 20.5K. To better understand the effectiveness of such c-axis pinning, the angular dependence of $J_c$ has been measured (right panel, Figure 7). At all fields, $J_c(\theta)$ exhibits a strong and broad *c*-axis peak. In particular, at low fields, the nanorods affect the whole angular range, significantly flattening $J_c(\theta)$. With increasing field, the c-axis peak still remains pronounced but in a smaller range around 90° and the *ab*-plane peak due to the mass anisotropy then emerges [77].

These results show that Ba-122 has remarkable tunable properties as a consequence of its high upper critical field, which allows a significant increase of the irreversibility field by pinning engineering, and its ability to accept a high density of non-superconducting secondary phases that act as strong pinning centers without detriment to the matrix superconducting properties. Of particular interest and value for applications are the low anisotropy ($\gamma$ <2), the high value of even the lower of the two upper critical fields ($H_{c2}^{\perp}(0) > 50$ T) and intra-grain $J_c$ values exceeding $10^5$ Acm$^{-2}$ over a very wide range. Sadly, as will soon be seen, there is clear evidence that GBs both in polycrystals and in bicrystals depress the intergranular current density $J_{cgb}$ significantly.

## II.4 Experimental methods for determining intra-and inter-grain J$_c$

In the case of a single grain boundary of a bicrystal thin film separating, separation of the grain boundary $J_{cgb}$ and in-grain $J_{cg}$ critical current densities is a straightforward matter of lithographically defining current tracks within the grain or across the GB and then determining $J_c$ using standard transport techniques. However it is not usually practical to isolate a single GB in bulk samples. Consequently, irrespective of whether the measurement is by transport, bulk magnetization, magneto-optical (MO) imaging or some other method, thought needs to be given as to how to separate the inter-grain and intra-grain contributions to the observed behavior.

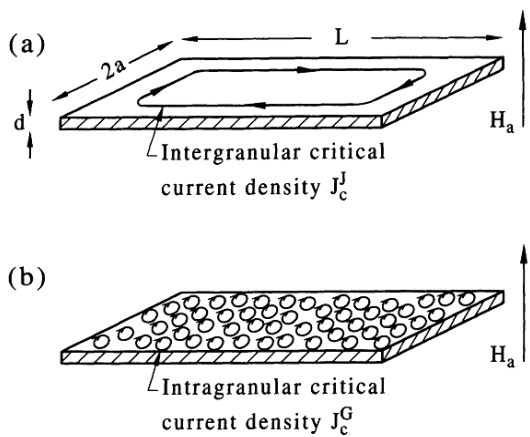

Figure 8. Comparison of inter- and intra-granular current flow in a polycrystalline material (from [81], Copyright 1994 by the American Physical Society)

Bulks are generally characterized using magnetization techniques rather than transport measurements since, even when $J_c$ is low, their size means that $I_c$ can be high. Determining grain boundary properties is complex, since the observed behavior is that of an assembly of a large number of grains and grain boundaries. Nevertheless there exist well established techniques for determining the length scales over which current loops flow from magnetization measurements. In a material in which grain boundaries are not obstructive, it is expected that current will flow in a loop encompassing the whole sample. However if current is restricted to individual crystallites a large number of small current loops are formed. Care, however, should be exercised, since many extrinsic factors like voids, inclusions, cracks and grain boundary wetting can give rise to poor connectivity and some of these have been observed in pnictide bulks [82]. Indeed while MgB$_2$ appears to have intrinsically benign grain boundaries, the preparation of fully connected bulks is challenging since GB obstructive phases and cracks can occur and high connectivity is difficult to obtain, Hot Isostatic Pressing is often required to produce the strongly connected bulk MgB$_2$.

It is possible to use the remnant magnetization method described by Müller *et al.* [81] to study the inter- and intra-granular currents flowing in polycrystalline bulks [68,82]. In this technique the measured moment is explicitly assumed to consist of two parts, one arising from intergranular current and the other from summing over all the intragrain loops. By measuring the sample whole and then after crushing it into progressively smaller pieces, it may be possible to determine the length scale on which the intra-grain currents are flowing by analyzing their contribution to the total moment. To quantitatively analyze the $J_c$ of two current loops requires knowledge of the length scales of the sample and the grains (or other scale of the current flow), which may become accessible if the sample can be ground to sizes where the smaller scale and the current flow length

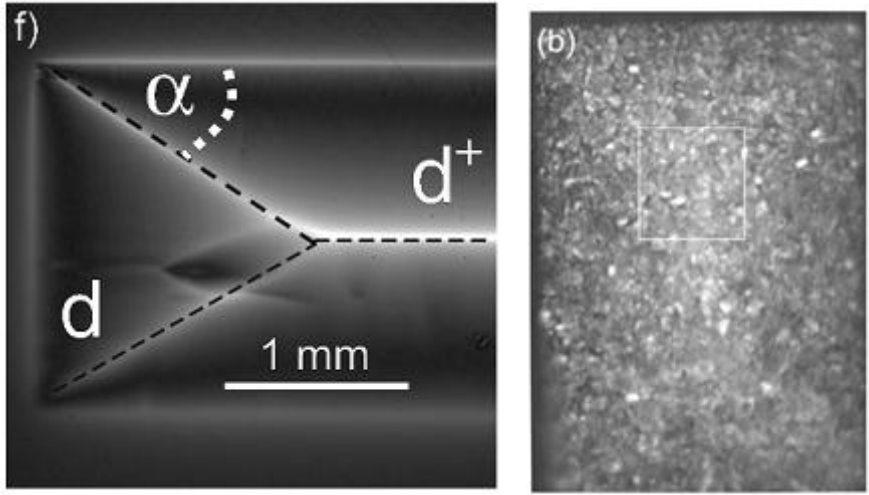

Figure 9 (a) A MO image showing at left the rooftop pattern corresponding to rather uniform long range current flow (Reprinted with permission from [84]. Copyright 2003, American Institute of Physics). It may be contrasted with the highly granular, weak-linked, current flow seen in the right hand sample (Reprinted from [85]).

are comparable. To observe the relative roles of the two scales of magnetic moment, a sample is exposed to many cycles of ever increasing magnetic field $H_a$, followed by removal of the field and measurement of the remnant moment, $m_R$. For a homogeneous sample, we can expect flux to penetrate when $H_a/(1-D)$ first exceeds the lower critical field $H_{c1}$, where $D$ is the relevant demagnetizing factor. For weakly coupled polycrystals, where the weakness could occur either at grain boundaries or at non-superconducting second phases, local flux penetration is expected to occur at lower fields than in the grains. Therefore we can separately observe the flux penetration transitions of inter- and intra-granular loops and $J_c$ values can be deduced by considering the sample and grain sizes and their respective geometrical factors.

Another way to evaluate the length scales of current flow is by using magneto optical (MO) imaging [83]. In a homogenous material, a clean and distinctive pattern is observed. For example, a rectangular shape generates a so-called "roof-top-pattern" (Figure 9b). Conversely, where current flow is confined to locally small current loops, a granular pattern appears, as in Figure 9a. If a sample shows at least a quasi-uniform bulk-scale critical state, the intergranular $J_c$ can be deduced from the local flux profile $b(x)$ using its gradient, $J_c = dB/dx$.

Another very powerful technique for evaluating the uniformity of current flow is low temperature laser scanning microscopy (LTLSM). The LTLSM allows a thin film sample to be examined below $T_c$ while biased with transport current sufficiently large that a measurable voltage occurs across the sample. A laser beam is scanned over the sample surface [86] [87] [88] producing a 2–3 μm diameter hotspot where the local temperature is increased. A local decrease in $I_c$ around the hotspot results in a small increase in voltage $dV$ across the whole sample if the $I_c$ of locally heated cross-section becomes smaller than the $I_{bias}$. The $dV$ response of the LTLSM measured at each point is proportional to the local electric field. [89] In non-uniform superconductors, a local $dV$ response is larger near blocking defects due to their preferential focusing of the supercurrent into channels [89] [69] LTLSM under an applied field provides information about the field-sensitivity of individual local current paths. Since Josephson-type weak-link paths across GBs are sensitive to fields, the LTLSM can often distinguish them rather clearly.

# III Experimental measurements of FBS grain boundaries

In this section we discuss recent results that address the properties of grain boundaries in various FBS systems. We start with the most straightforward method of measuring the critical current of a grain boundary using a thin film of known misorientation θ grown epitaxially on a bi-crystal substrate. We then discuss results obtained by separating inter-grain and intra-grain properties in polycrystals and wires using some of the techniques described in the last section.

## III.1 Thin films

A proven technique for studying individual grain boundaries with different misorientations θ is to grow a thin film on a bicrystal substrate that allows epitaxial overgrowth of the superconductor, as extensively employed for the study of grain boundaries in the cuprate superconductors. Given the uncertainties involved in decoding intrinsic GB properties from polycrystals, it is unsurprising that this technique is also very desirable for FBS superconductors too. However, epitaxial thin films of FBS superconductors have been hard to grow, especially the F-doped rare earth (RE) 1111 phase. Because both As and F are volatile at the deposition temperature, it is difficult to control the composition of the deposited film, especially the F:O ratio which controls the carrier density. By contrast, Co, which can be used as the dopant in $Ba(Fe,Co)_2As_2$ (Ba122), has a low vapor pressure under growth conditions and the most extensive study so far has been made with this 122 compound [58,59]. Nevertheless, it is still not easy to grow epitaxial films of the FBS. One fundamental difficulty is the structural and chemical mismatch between the metallic FBS compound and the usual single crystal oxide substrates. Lee *et al.* [79] have overcome this by employing an epitaxial template layer of a divalent, alkaline earth (AE)-containing $SrTiO_3$ (STO) or $BaTiO_3$ (BTO) between the single-crystal perovskite substrate and the Ba-122 film so that they have a common interfacial layer. In this way, it has been possible to grow epitaxial Co-doped Ba-122 thin films of very high crystal perfection on both (001) oriented STO bicrystal substrates and (001) oriented LSAT bicrystal substrates using an intervening $SrTiO_3$ (STO) template layer [58]. Epitaxial films have also been grown by the Dresden and Tokyo groups without such template layers by carefully adjusting the growth conditions [90],[91].

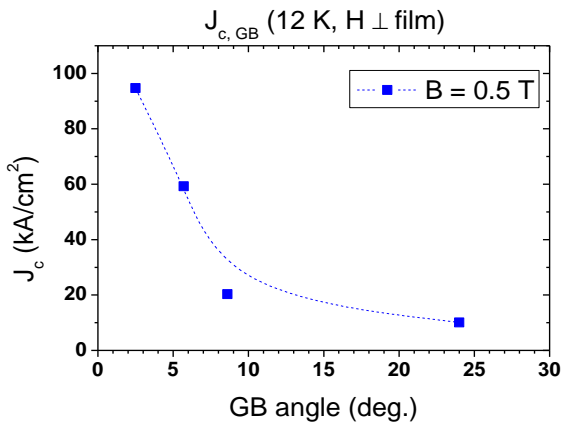

**Figure 10.** Variation of GB critical current density $J_{cgb}$ with [001] tilt misorientation angle θ for 8%Co-doped Ba-122 thin film bicrystals [58].

The intragrain $J_c$ of such Ba122 films is over 1 MA/cm$^2$ (4.2K, SF) which has allowed Lee *et al.* to study the GB weak-link behavior as a function of GB misorientation θ [58]. Their essential conclusion is that [001] tilt grain boundaries in FBS begin to depress $J_{cgb}$ below $J_{cg}$, as in YBCO, for θ >3°. With only 4 different values of θ, the authors could not confirm an exponential relationship between θ and $J_{cgb}$ as in YBCO, though Figure 10 does show a remarkable similarity to the exponential fall of $J_c(\theta)$ found in the cuprate superconductors [15]. They did note however the 9° [001] tilt GB exhibited less depression of the $H_{irr}$ than did similarly misoriented YBCO bicrystals, a result suggesting a more metallic GB in the FBS.

Katase *et al.* [59], performed a similar study, also on 8%Co-doped Ba-122 and additionally they looked at the properties of Josephson junctions made with higher angle bicrystals. Contrary to the view of Lee *et al.* that the critical angle was ~ 3°, they proposed that the critical angle for [001] tilt boundaries was ~ 9°. They also studied the properties of the Josephson-coupled grain boundaries and concluded that they showed SNS rather than SIS behavior, a result which points to higher angle GBs being metallic rather than insulating, as in the cuprates. This result is consistent with the conclusion of Lee *et al.* based on the behavior of $H_{irr}$.

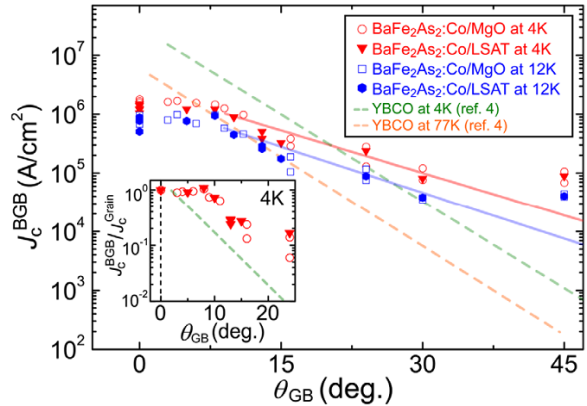

**Figure 11 Variation of intergranular $J_c$ with grain boundary misorientation angle θ [59].**

While at first sight the conclusions of Lee *et al*. and Katase *et al.* data sets appear to disagree, we believe that the differences are minor. Both studies show a similar decrease in $J_c$ with misorientation angle that is consistent with an exponential fall off of $J_{cgb}$ with increasing θ. The issue dividing the two studies is where the critical angle at which $J_c$ becomes limited by the GB occurs. Is it as low as 3° or as high as 9°? The differences may well turn out to be related to subtleties of GB structure, differences in the magnitude of the $J_c$ in the grains on either side of the GB, or other as yet uncontrolled parameters. This difference in $θ_c$ may not be surprising, given the similarities of FBS and cuprates, where considerable scatter in $J_{cgb}$ values on GBs of the same misorientation angle has been well documented [15].

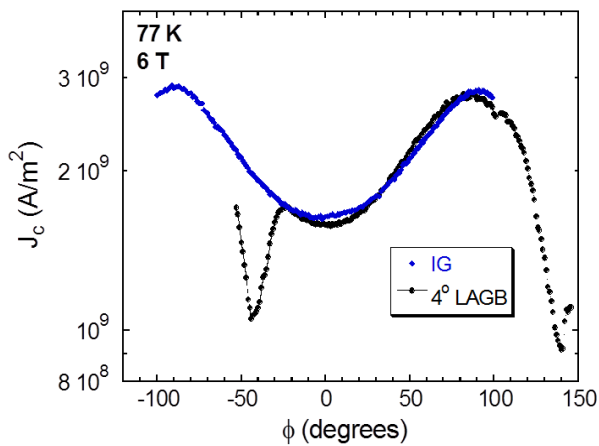

**Figure 12 Comparison of $J_c$ with in-plane field rotation ϕ (see Figure 13) for a intragrain track (IG) and a track crossing a 4° [001] tilt grain boundary in an epitaxial thin film YBCO bicrystal (B lies //GB-plane at ϕ=–45°).**

An alternate approach to extracting the way that the GB properties affects the depinning of vortices and $J_{cgb}$ and $J_{cg}$ of the whole film can be obtained from angular-dependent studies made with the magnetic field **B** lying in the plane of the film. In the absence of any GB, $J_c$ varies sinusoidally as **I** and **B** vary from being orthogonal (ϕ = 0°) where there is a full Lorentz force and $J_c$ is minimum to parallel and $J_c$ is maximum (ϕ = 90°). If the GB has properties different from the grain, this geometry is very valuable in showing them up. Durrell *et al.* have explored this situation using both YBCO bicrystals [43] and Co-doped 122 bicrystal films [92] made by Lee *et al.* [58]. Figure 12 shows a typical result from a 4° [001] tilt YBCO bicrystal, where, over most values of misalignment angle $\phi$, there is no distinction between $J_{cgb}$ and $J_{cg}$. Only when **B** is near to or parallel to the GB does $J_{cgb}$ drop because only then does an appreciable vortex length lie in the GB, where it can become A-J like if the GB has depressed superconducting properties. This is indeed the situation for this 4° [001] tilt GB, just 1° above the usually quoted minimum value of $\theta_c$ = 3°. The critical event triggered by this vortex geometry is that the more weakly pinned A-J vortex segments lying in the GB depin at a lower Lorentz force than do the intragrain Abrikosov vortices. When only very short segments of vortex cross the GB, the overall vortex line tension and pinning forces exerted on the intragrain segments are strong enough to prevent depinning of the GB vortex segment until depinning of the intragrain segments occurs.

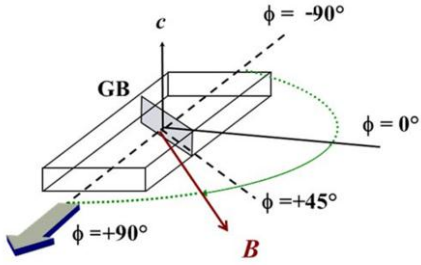

**Figure 13** Geometry of the measurement scheme used in Figures 12, 14 and 15. $J_{cg}$ and $J_{cgb}$ are measured as the angle $\phi$ is varied (B//I with $\phi$=90°, B⊥I with $\phi$=0°)(42). Here the track is patterned at 45° to I and to the GB plane, although this angle may be arbitrarily chosen when the current track is lithographically defined. The track was patterned at -45° for the YBCO work in Figure 12 and at 45° for the Co-doped 122 films of Figure 14 and 15.

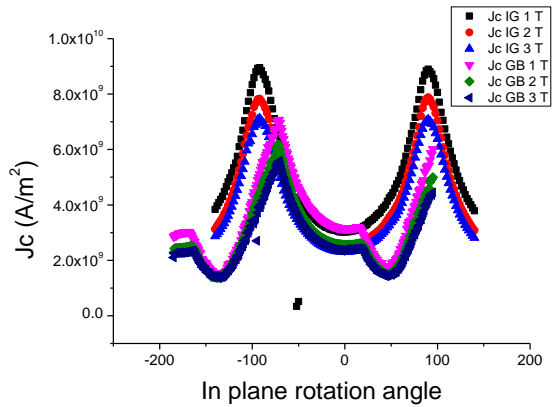

**Figure 14** Vortex channeling in a Co-doped Ba-122, 5° [001] tilt bicrystal at 5K. The upper periodic curves show the expected variation of the intragrain (IG) $J_c$ as B rotates from the force-free condition (I parallel to B at -90° and +90°) to the full Lorentz force condition (B perpendicular to I). The lower curves show the intergranular $J_c$ (see the geometry of the GB in Figure 13). Over much of the angular range $J_{cg}$ and $J_{cgb}$ follow, except where B and the GB plane nearly coincide. $J_{cgb}$ falls sharply in this region when an appreciable length of vortex lies in the GB plane.

In contrast to the study of Durrell *et al*. [43] on YBCO bicrystals, where the GB was aligned perpendicular to the track and to current flow, the tracks in the Co-doped 122 GBs were patterned so that the GB was at 45° to the current direction so as to reduce the symmetry of the measurement. Figure 14 shows that $J_{cgb}$ and $J_{cg}$ are coincident between about -70° and +20° for a 5°[001] tilt Co-doped 122 GB. Figure 14 also shows that near $\phi$ = 45° and -135° there is a significant dip in $J_{cgb}$ where a significant vortex kink lies in the GB. Over all other angular ranges, the GB vortex kink straightens out and the vortex lies parallel to the field vector, such that only a very short length of vortex lies in the GB and vortex depinning switches from being determined by the weaker-pinning GB to the stronger pinning grains.

On moving from a 5° to a 12° [001] tilt bicrystal (compare Figure 14 and 15), the degradation of the GB properties at higher misorientation angle $\theta$ is clear. For the higher angle GB, there is no value of $\phi$ at which $J_{cgb}$ ever reaches $J_{cg}$, leading us to conclude that the whole GB has significantly depressed superconducting properties. Even increasing the doping element Co from 8 to 10% fails to improve matters.

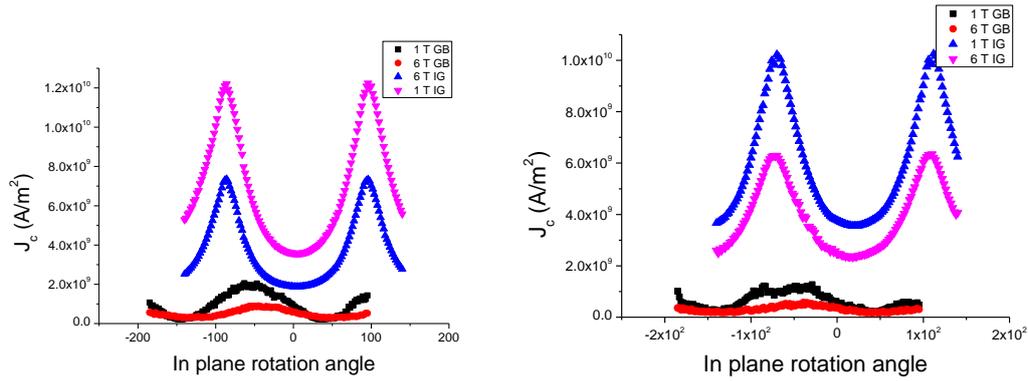

Figure 15 Vortex channeling in 12° [001] tilt Co-doped Ba-122 bicrystals at 5K (a) 8% Co doping (optimal doping) and (b) 10% Co doping (over doping). Unlike the data shown for the 5°[001] tilt bicrystal of Figure 14, the $J_c$ measured across the GB is here always lower than the intragrain $J_c$ for all values of ϕ. The overdoped sample has even lower $J_{cgb}$ than does the optimally doped sample.

In summary, we conclude that all three thin film bicrystal studies performed so far on FBS come to qualitatively the same conclusion. At some misorientation less than 10°, [001] tilt GBs exhibit signs of reduced superconducting properties, most specifically the situation where $J_{cgb}$ becomes less than $J_{cg}$. A first attempt at overdoping has not shown any improvement of $J_{cgb}$, unlike the situation for Ca doping of YBCO where adding carriers to the compound does improve $J_{cgb}$ [35,44,46]. Clearly many more such experiments are desirable given the very attractive properties for applications that even this 22 K $T_c$ superconductor possesses. One can quite explicitly say that only the poor $J_c$ of random polycrystals holds back applications, as the next sections now discuss.

## III.2 Polycrystals

Compared to the bicrystal films discussed in the previous section the inter-granular $J_c$ in randomly oriented polycrystalline samples is more complicated. In a polycrystal, millions of crystallites with varying chemical compositions, sizes, shapes and orientations form a network through which current must percolate. This percolation path must cross grain-boundaries with varied misorientation angles and defects. Deduction from macroscopic electromagnetic measurements, supported by direct observation with local information is an important approach to assess and comprehensively understand the inter- and intra-granular current behaviors in FBS polycrystals. Such studies remain of great importance whatever the uncertainties of the sample quality because as noted in the previous section, thin film bicrystals have only been made so far of the Co-doped 122 FBS compound. Here therefore we concentrate on studies of the highest $T_c$ family of FBS, those based on the RE1111 compounds.

When we examined our first La-1111 polycrystal samples (made at Oak Ridge National Laboratory), we found quasi-reversible magnetization hysteresis loops suggesting either widely spread obstacles to inter-granular current or very weak vortex pinning. Study of the magnetization in bulk and powdered samples showed almost no sample-size dependence of magnetization, indicating substantial electromagnetic granularity on a scale approximating the grain size [93]. These samples were substantially less than 100% dense, thus introducing many voids as current-blockers, a point not always properly recognized by studies of sintered complex compounds.

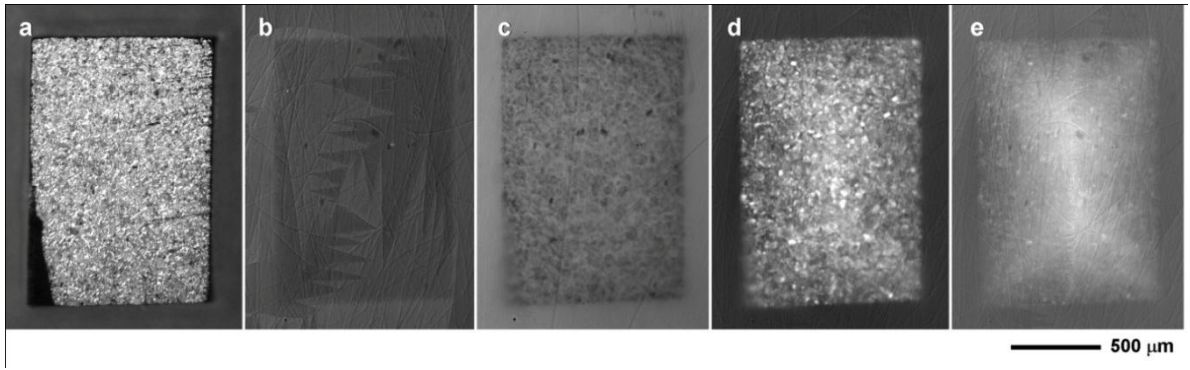

**Figure 16.** Reflected light and magneto-optical (MO) images taken on a dense SmFeAsO$_{0.85}$ bulk sample. (a) Light microscopy image of the polished surface. (b) and (c) MO images of different stages of magnetic flux penetration into the sample when zero field cooled (ZFC) to 6 K after a perpendicular field of 4 mT (b) and 120 mT (c) was applied. (d) and (e) MO images in zero field at 6.4 K (d) and 20 K (e) after field cooling at 120 mT (Reprinted from [68]).

Polycrystalline Sm and Nd 1111 samples made by high pressure synthesis at Chinese Academy of Science were much denser and showed significantly enhanced hysteretic magnetization [68]. Indeed relatively high inter-granular $J_c$ of the order of 1000 A/cm$^2$ in randomly oriented polycrystals was a big surprise for us. Figure 16 shows magneto-optical (MO) images taken on a well-polished surface of this Sm1111 bulk sample. Under the zero-field-cooling (ZFC) condition, we observed a bulk Meissner state when a field of 4 mT was applied at 6 K, as shown in Figure 16b, which indicates that the surface shielding current flows over the whole sample. As the external field increased, flux started to penetrate at ~6 mT and first reached the center of the sample at ~15 mT (Figure 16c). However the MO images clearly show that flux penetration is quite inhomogeneous. There are many 20-50 µm size black-appearing spots of strong flux shielding in the flux penetrated regions, indicative of local circulating currents with higher current density than the average of the whole. Figure 16d and e show MO images under perpendicular fields of 0 mT at 6.4 and 20 K, respectively, after cooling in 120 mT. It is particularly noteworthy that the strongly coupled local regions were clearly visible at 6.4 K and that raising the temperature to 20 K produced a quasi-ideal, roof-top pattern derived from a global, whole-sample current that largely washes out the less visible granular structure.

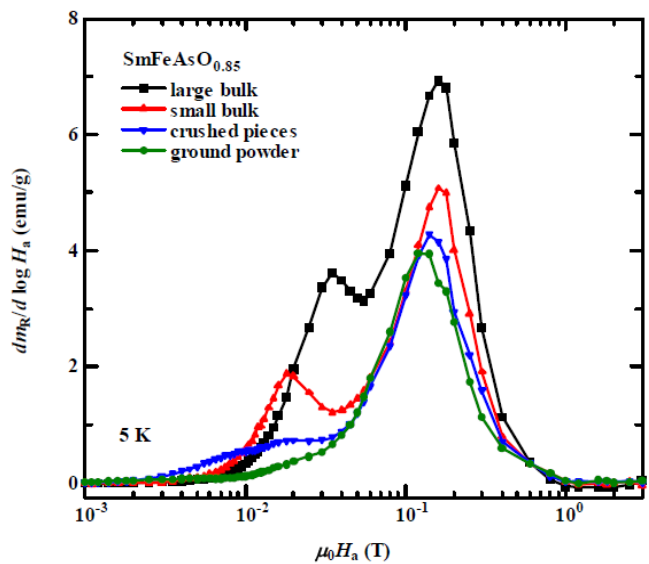

**Figure 17.** Derivatives of the remnant magnetization ($m_R$) as a function of increasing applied field at 5 K for an intact piece of the SmFeAsO$_{0.85}$ large bulk, intact a small bulk, crushed pieces and ground powder derived from one original sample. The data are normalized by the sample masses of 30.0, 6.0, 14.2 and 4.3 mg, respectively (Reprinted from [68]).

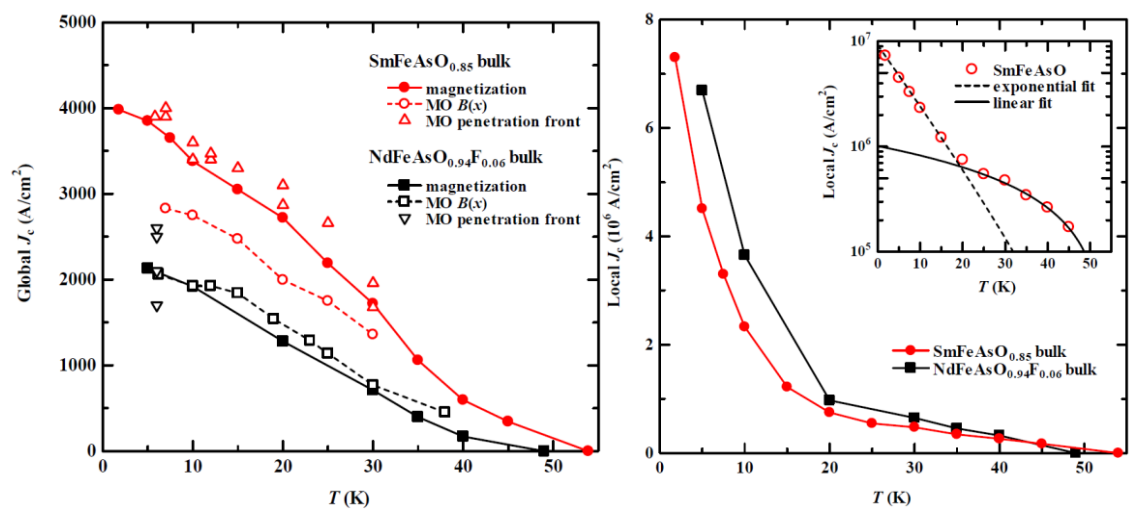

Figure 18. (a) Temperature dependence of inter-granular critical current density for SmFeAsO$_{0.85}$ and NdFeAsO$_{0.94}$F$_{0.06}$ bulk samples obtained from the remnant magnetization analysis (filled symbols), magneto optical B(x) flux profile analysis for the FC conditions, and the width of the flux penetration front in the ZFC conditions using Brandt's expression [94] for a strip in perpendicular field. (b) Temperature dependence of critical current density of locally circulating current $J_{cg}(T)$ for SmFeAsO$_{0.85}$ and NdFeAsO$_{0.94}$F$_{0.06}$ bulk samples obtained from remnant magnetization analysis. Inset shows log-scale plots for the SmFeAsO$_{0.85}$ data with exponential and linear fits to $J_c(T)$ (Reprinted from [68]).

In order to make a more explicit test of the scale over which currents flow, we made remnant magnetization analysis on polycrystalline samples of different size. Figure 17 shows the derivative of the remnant moment $m_R$ as a function of increasing applied field for the Sm1111 intact large bulk, a second, smaller bulk, relatively large crushed pieces and finely ground powder. For the bulk samples, $m_R$ began to increase on raising the applied field above ~5 mT. Two quite separate low-field peaks appear, at 35 mT for the larger and at 18 mT for the smaller of the two bulk samples. By contrast, the second, higher field peak appeared at the same field of 150 mT for both samples. That the first peak strongly depends on sample size indicates that the bulk current loop is size dependent, because the field of first penetration should be proportional to $H_p \sim J_c^{global} \times$(sample size). This conclusion is strengthened by further suppression of the first peak in the crushed pieces and its disappearance almost to zero in the ground powder. In contrast, the second peak was found to be size independent, even after fine powdering, which means that the second peak is caused by locally circulating currents with current loop size less than the powder size of 20-50 μm.

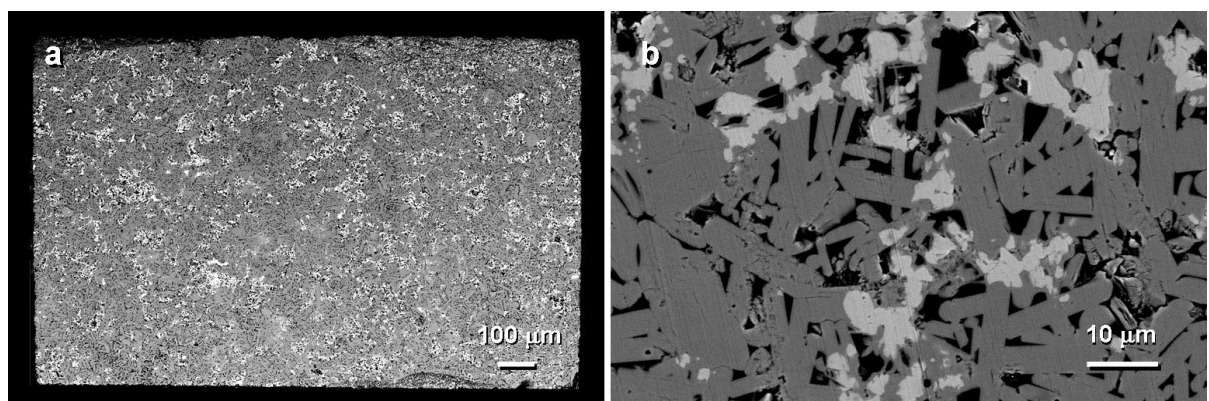

Figure 19. Scanning electron images of the polished surface of SmFeAsO$_{0.85}$ showing a dense, but still multi-phase microstructure. Although this is an example of a good bulk sample of a polycrystalline FBS, there are still many extrinsic obstacles to current flow in such samples (Reprinted from [68]).

Figure 18 shows the temperature dependence of the inter- and intra-granular $J_c$ (these are of course approximations to the quantities $J_{cgb}$ and $J_{cg}$ that can be explicitly measured in a bicrystal experiment) deduced from the lower and higher peaks of the $dm_R/d\log H_a$. The inter-granular $J_c$ was also calculated from the MO flux profiles and the width of the flux penetration front in the ZFC images using formulae in references [83,94,96]. The agreement between these various methods of extracting the inter-granular $J_c$ is excellent. The temperature dependence of the inter-granular $J_c$ is almost linear, except near $T_c$, and $J_c$ values at 5 K of 3900 and 2100 A/cm$^2$ for the Sm- and Nd-1111 samples, respectively, were obtained. The temperature dependence of the intra-granular $J_c$, for currents circulating only within individual grains, is shown in Figure 18b. The $J_c$ values are 4.5×10$^6$ and 6.7×10$^6$ A/cm$^2$ at 5 K for the Sm- and Nd-1111 samples, respectively. Both samples showed strong upward curvature in $J_c(T)$ below ~20 K.

Figure 19a shows a whole sample image of the Sm1111 bulk obtained by scanning electron microscopy. High magnification images of the polished surface of the Sm1111 revealed plate-like grains of the superconducting phase with a size of ~20 μm, as shown in Figure 19b. The sample contains multiple impurity phases, with the two most prominent phases, as shown in Figure 19b, identified by EDS as $Sm_2O_3$ (white contrast) and Fe-As glassy phase (dark contrast). It appears that the microstructure is rather homogeneous on the macro-scale, although the Nd bulk sample showed macro-scale inhomogeneity on a scale of several hundred micrometers [85].

Multiphase microstructures are so far a common problem among all the types of FBS polycrystals (as indeed is still generally true for cuprates). Here we show an example of microstructure of one of the most complicated types of FBS, $Sr_2VO_3FeAs$ [97] with a perovskite block layer. As-prepared $Sr_2VO_3FeAs$ bulk was synthesized from FeAs, Sr, SrO and $V_2O_5$ powders and then crushed into powders, pressed into a pellet and sintered at 1200°C for 24 h (sintered bulk) at Univ. of Tokyo [95]. Figure 20 shows back scattered electron (BSE) images of a polished surface of the as-prepared and sintered $Sr_2VO_3FeAs$ bulks. The as-prepared sample shows a distinctly inhomogeneous structure with a length scale of several tens of micrometers consisting of $Sr_2VO_3FeAs$, $Sr_2VO_x$ and FeAs phases (Figure 20a). After sintering, the microstructure becomes denser and the impurity phases decrease; however cracks between grains appear and traces of the very common wetting FeAs phase remain, as shown in Figure 20b. We observed similar cracks and wetting FeAs in polycrystals of even our best Re-1111 and Co-doped 122 samples too [98].

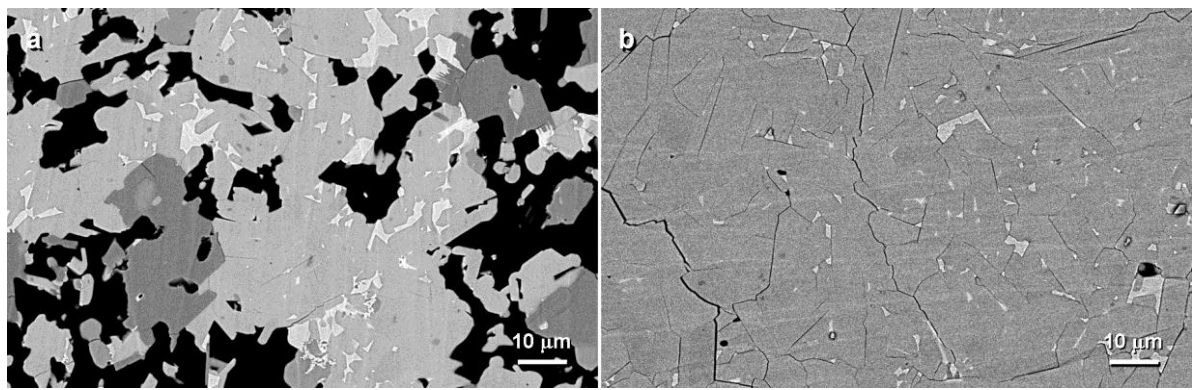

Figure 20. Back scattered electron images of a polished surface of the as-prepared and the sintered $Sr_2VO_3FeAs$ polycrystals. Light gray, dark gray, white and black contrasts correspond to $Sr_2VO_3FeAs$ superconducting phase, $Sr_2VO_x$ phase, FeAs wetting phase and pores or cracks, respectively [95].

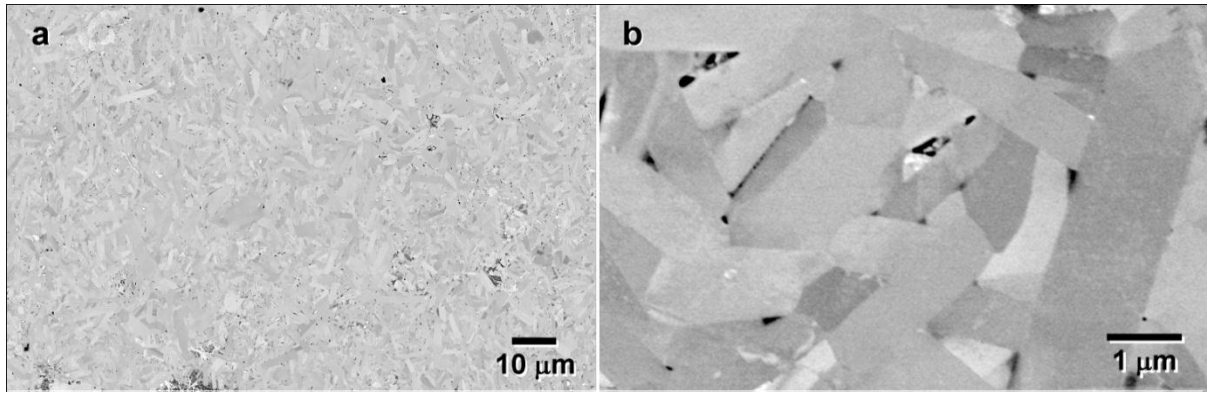

Figure 21. Low (a) and high (b) magnification secondary electron microscopy images for the SmFeAsOF$_{0.85}$ bulk hot-isostatic-pressing (HIP) processed at 900°C under 200 MPa (Reprinted from [93]). These samples have significantly improved phase purity compared to the sample shown in Figure 20, but actually have lower $T_c$ due to F loss and lower intergranular $J_c$ values.

The challenge of improving the phase purity of polycrystalline Sm1111 bulk samples is significant, as we demonstrate by one recent set of experiments in which we employed sintering and hot isostatic pressing (HIP) [93]. An as-prepared SmFeAsO$_{0.85}$F$_{0.15}$ sample was synthesized by solid-state reaction at low pressure from Sm, As, Fe, Fe$_2$O$_3$ and FeF$_2$ at the University of Genova. The as-prepared sample was ground into powder, pressed into a pellet, and then either sintered at 1250°C for 24 h in an evacuated, sealed quartz tube, or HIP processed at 900-1200°C for 10 h under 200 MPa. The microstructure of the as-prepared Sm1111 bulk, with a relative density of 50-60%, has areas of dense Sm1111 grains but these are separated by large pores. The wetting and normal-conducting FeAs phase is present between many Sm1111 grains, even in the dense regions, similar to Figure 21(b). As the processing proceeds from sintering to HIP processing, the samples become denser, increasing from ~75% in the sintered bulk to ~90% in the HIP processed bulk. The microstructure becomes very dense after HIP processing as shown in Figure 21. Here, all the dark regions are impurity phases, not pores. In the HIP processed sample, we did not observe the wetting phase or a continuous network of cracks (Figure 21a). HIP processing almost eliminates the grain boundary wetting phase, as well as densifying the sample, which is clearly desirable. However, in spite of this careful and complex processing, the increase in the intergrain $J_c$ was quite small. We believe that the microstructure of the HIP processed samples are some of the best reported in the literature to date from the view point of phase purity and inter-grain connections, since many 1111 polycrystalline literature samples contain macroscopic impurity phases and/or wetting phases at grain boundaries. Evaluation of the balance between a so-called intrinsic property (GBs have a fundamental tendency to develop depressed superconductivity as discussed in section II.2) or these impurity phases, cracks, voids etc. is seldom easy. In this case, the enhanced phase purity and density was bought at the price of lowered $T_c$ due to fluorine loss. Given that the highest global or intragranular $J_c$ of ~4000 A/cm$^2$ was obtained in an earlier dense CAS sample of Sm1111 with $T_c$ of ~55 K, while these HIP processed samples showed $T_c$ of ~37-45 K, the lowered $T_c$ could be a crucial factor affecting the intergrain $J_c$ since $T_c$ is correlated with doping and thus with carrier density. Until it becomes possible to grow high-quality epitaxial films of the 1111 compounds, this frustrating state of affairs is likely to remain

## III.3 Wires

Since any magnet technology depends on wires kilometers long, the issue of whether current can traverse grain boundaries present in manufactured wires is of great practical importance. While magnet applications of FBS are not yet being actively considered, the FBS do have the required $T_c$, $H_{c2}$, intragranular $J_c$ and low anisotropy to be interesting for applications. The principal obstacle to applications is actually the low value of $J_c$ measured in polycrystals and the concern that grain boundaries are intrinsically weak linked at all except the smallest misorientations. As previous sections have shown, only one compound, Co-doped Ba-122, has yet allowed high quality thin film growth by multiple groups. Thus broad evaluations of the whole class of FBS compounds must for now be done with bulk forms. As noted in the previous section, experiments that reliably extract GB properties from bulks are not easy to devise or to carry out, and the understanding that they yield of $J_c$ properties is hard won. Wires may play an important additional role because the metallic sheath can, if chosen properly, encapsulate the FBS constituents and allow new synthesis approaches, as well as making transport measurements of $J_c$ much easier. Wires may thus have important scientific and technological roles to play in understanding the science of FBS.

The most common route to wire manufacture is through powder-in-tube (PIT) techniques in which powders are packed in a metal tube, the tube drawn into a wire, and then heat treated to form a connected superconductor at final size. If the FBS compound is synthesized before loading into the tube, this is named an *ex situ* process. *Ex situ* routes give more options for processing, because the constituents can be ground and reheated multiple times in pursuit of complete reaction. This is particularly important when working with rare earth, alkaline earth, and alkali metal elements that are soft and gummy and thus difficult to grind. Reaction inside the tube, that is by an *in situ* process, may be helpful when concerns about loss of material or reaction in open systems dominate. If inert, or at least only benignly reactive, tubes can be used to enclose the reaction, the potential benefits of fine grinding produced during the wire or tape deformation may appear. Extensive particle fracture should occur during drawing and rolling, perhaps making final reaction quicker and more uniform. In either case, connections to the superconductor through the metallic sheath will make transport current measurements easier than for most small bulk samples, for which, typically, only magnetization measurements are possible. Another advantage of a wire is that longer samples, perhaps 1 m rather than just a few mm, are generally possible, allowing access to much larger ensembles of grain boundaries. For multiple reasons therefore, continued study of wire forms of FBS has value.

In fact the first FBS wires were made shortly after FBS were discovered from the 1111 [64,99] and 122 [99] compounds. Examples of *in situ* powder mixtures are Sm, Fe, As, Fe, $Fe_2O_3$, and $SmF_3$ for Sm-1111 [64] and Ba, K, Fe, and As for K-doped Ba-122 [100]. The *in situ* approach is valuable in light of the toxicity, high vapor pressure and volatility of some of the reactants, especially As. The *in situ* PIT process is particularly advantageous for 1111 wire that has to be doped with F, which can easily vaporize during the heat treatment. The typical disadvantage of *in situ* PIT wire is that the only experimental parameters available to fine-tune the synthesis of the complex phase are the time and temperature of the final heat treatment. Because FBS phases are brittle, it is generally thought unwise to deform them after formation, which would rule out any use of further deformation to better mix residual unreacted constituents to a better reaction status. However Togano *et al.* have performed such experiments [100]. Indeed an intermediate rolling step between the two heat treatments is a key part of the optimum production process for Bi-2223 [13], making this step less unrealistic than might be thought at first sight. Another advantage of e*x situ* powders is that their quality can be much better assessed before they are packed into the tube, again allowing greater freedom to the FBS synthesis process.

The initially chosen metal tubes were Ta, or Nb, or Fe with an inner Ti foil. However, all these materials reacted with the FBS during heat treatment, which were typically performed at 1100 to 1200°C. Ag is now being used since it does not react with the FBS materials or affect their superconducting properties. However, the low melting temperature (961 °C) of Ag does restrict reaction temperatures to 850 to 900 C.

Figure 22 shows some of the highest reported transport $J_c$ data for FBS PIT wires. Although self-field 4.2 K values can reach as high as ~5000 A/cm$^2$, the application of much less than 1 T depresses $J_c$ by an order of magnitude to 100-1000 A/cm$^2$. These $J_c$ values are two to three orders of magnitude lower than for the single crystal films shown in Figure 7. Consistent with all the prior discussion, this is likely due to weak links that form at randomly misoriented grains, as well as SNS connections that form across GBs coated with the normal Fe-As wetting phase [85]. The weak $J_c(H)$ dependence at fields of more than a few tesla indicates that the grains themselves probably have intragrain $J_c$ properties characteristic of good thin film samples and the dense bulks evaluated by Yamamoto *et al.* [68] (Figure 18). The degraded connectivity of wires is probably amplified by residual cracks and porosity, but these are scientifically irrelevant to the present discussion. Figure 22d shows the cross section of a Ag-sheathed ex situ (Ba,K)Fe$_2$As$_2$ wire that has one of the highest $J_c$ performance reported to date, but it is obvious even from this low magnification image that the wire is far from having an ideal, dense, single-phase microstructure. There is thus still much room for improvement before we can conclude that such properties are "intrinsic". In this case $J_c$ reaches almost 10,000 A/cm$^2$ at self field and 1000 A/cm$^2$ at 15T at 4.2K. It seems clear that progress in raising the transport $J_c$ of wires is occurring and that PIT approaches may actually be amongst the most promising to apply for making bulks samples, even if one reason for the interest is as a convenient way to react the FBS compounds in an enclosed environment [99],[101].

*Ex situ* wires may offer a greater opportunity for improving $J_c$ because, as mentioned above, there are more options for remixing unreacted materials and reacting them multiple times to drive the reaction to completion. Weiss *et al.* [104] have developed a technique to make high-purity (Ba,K)Fe$_2$As$_2$ using a SPEX high-energy ball mill to mix the constituent elements. The high-energy milling initiates a self-sustaining reaction that forms a brittle Ba-122 powder which fractures in the mill, allowing better mixing and a more complete reaction. Bulk samples made from a reaction of this powder at 600 °C were essentially phase pure, and had a whole-sample $J_c$ calculated from magneto-optic images of ~50 kA/cm$^2$ at 5.7 K and self-field. This is the highest whole-sample $J_c$ yet reported for any FBS bulk, suggesting that control of many extrinsic defects is still valuable. Such values of $J_c$ begin to signal that applications of FBS might be possible since they start to become compelling once $J_c$ values greater than $10^5$ A/cm$^2$ become feasible in an interesting domain of B and T. Another route being explored to raise $J_c$ is by enhancing the connectivity of the FBS core structure by adding metallic elements to the FBS. Ag and Pb powder have been added to improve the grain connectivity and fill cracks and pores in PIT wires [66] [103] [101] too. Figure 22c shows that these additions can increase the magnitude of the self-field $J_c$ and the $J_c$ plateau at high field, but also that they do not eliminate the large low-field weak link drop of $J_c$.

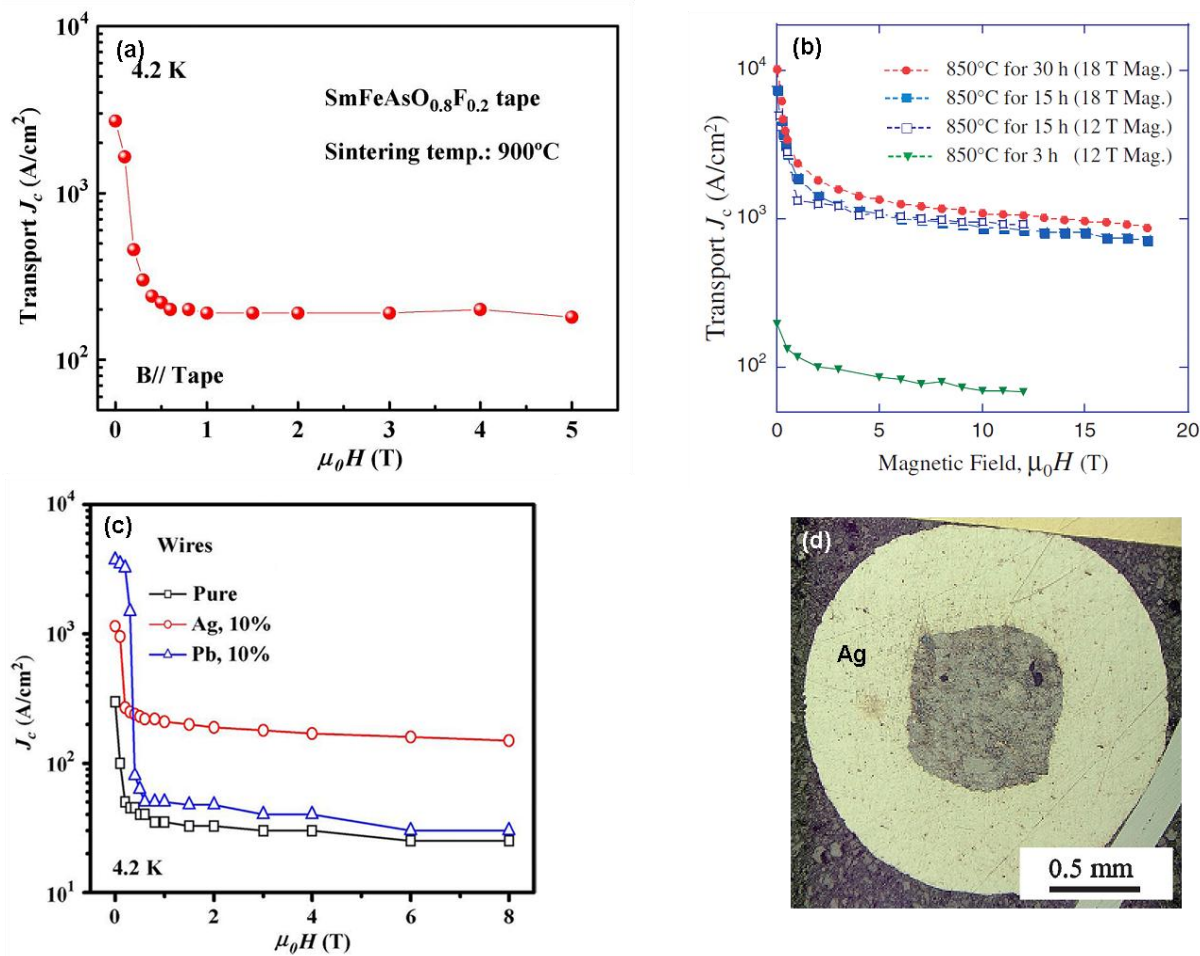

Figure 22. Transport $J_c$ at 4.2 K for Ag-sheathed (a) *in situ* SmFeAs(O$_{0.8}$F$_{0.2}$) rolled tape [102] (© 2011 IEEE), (b) *ex situ* (Ba$_{0.6}$K$_{0.4}$)Fe$_2$As$_2$ with 13 wt% Ag added [103] (Copyright 2011 The Japan Society of Applied Physics), and (c) transport $J_c$ as a function of applied field for pure, 10 wt% Ag-, and 10 wt% Pb-doped (Sr$_{0.6}$K$_{0.4}$)Fe$_2$As$_2$ wires [102] (© 2011 IEEE). The Sm-1111 tape in (a) was rolled from a round wire. (d) Light microscope image of the cross section of the 850 °C/30 h wire in (b) [103] (Copyright 2011 The Japan Society of Applied Physics).

The generally low $J_c$ values of prototype wires and the evidence for weak link behavior from the Co-doped thin film bicrystals have naturally generated interest in growing FBS films on the textured templates used for YBa$_2$Cu$_3$O$_{7-x}$ coated conductors [14]. Recently Co-doped Ba-122 coated conductors have been grown on textured MgO buffer layers on Hastelloy substrates. Iida *et al.* [90] deposited a 50 nm Fe layer (Fe seems to perform the same function as the SrTiO$_3$ template used by Lee *et al.*[79]) on the MgO and then used PLD to deposit the Co-doped Ba-122 film on the Fe layer. Katase *et al.* [91] used PLD to deposit the Co-doped Ba-122 layer directly on the MgO. Figure 23 shows that the transport $J_c$ for these Ba-122 coated conductors reaches ~1.5 MA/cm$^2$ at 4.2 K, SF. The FWHM of the in-plane misorientation distribution $\Delta\phi$(Ba-122) in the Iida *et al.* film was 5° and ~3.2-3.5° in the Katase *et al.* films. Katase *et al.* noted that the Ba-122 in-plane alignment was improved compared to that seen in the textured MgO layer: $\Delta\phi$(MgO) was 7.3°, 6.1°, and 5.5° while the $\Delta\phi$(Ba-122) grown on these MgO layers was 3.2° to 3.5°. All of these $\Delta\phi$ (Ba-122) values are within the grain-to-grain misorientation range where we expect the GB to offer minimum obstruction to transport current [58] [59], as discussed in section III.1. This coated conductor template approach has also been used for the deposition of the As-free Fe(Se,Te) 11 compounds and similarly high $J_c$ values have been obtained [105]. Implicitly of course, this route to higher $J_c$ by epitaxial growth on a highly textured substrate confirms the conclusion from the thin film bicrystals that the weak link problem is intrinsic to FBS.

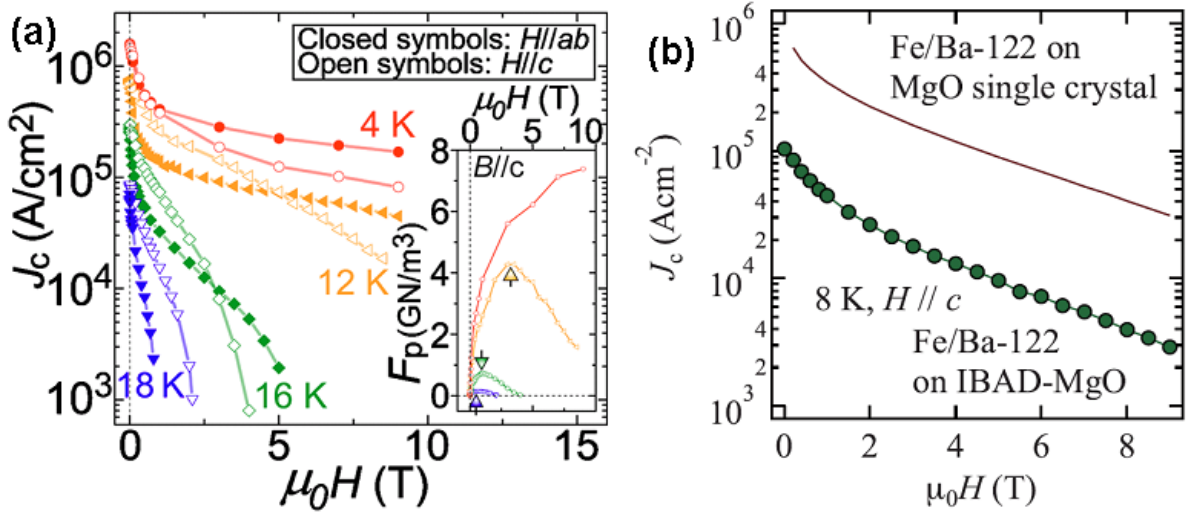

Figure 23. $J_c$ as a function of applied magnetic field for a Co-doped Ba-122 film deposited on an IBAD- textured MgO template layer grown on a Hastelloy substrate. Ba-122 film deposited (a) directly on the MgO (Reprinted from [91]. Copyright 2011, American Institute of Physics) and (b) on a ~10 nm thick layer of Fe deposited on the MgO (Copyright 2011 The Japan Society of Applied Physics) [90]. The inset in (a) shows the pinning force $F_p$ for B//c at 4.2 K. $J_c(H)$ for a Ba-122 film deposited on a thin Fe layer deposited on a single crystal of MgO are several times higher as is shown in (b).

Interestingly, the transport $J_c$ values increased with increasing Δϕ (MgO) achieving 3.6, 1.6, and 1.2 MA/cm$^2$ at 2 K SF on films with Δϕ (MgO) = 7.3°, 6.1°, and 5.5°, respectively. This is exactly the opposite effect expected and probably indicates that other issues are present in these early films. Both reports note that the $J_c(H)$ behavior indicate strong c-axis pinning, as is also evident in the data of Fig. 7 for Co-doped BaFe$_2$As$_2$ films grown epitaxially on single-crystal oxide substrates [77].

In summary therefore we can say that transport $J_c$ values of PIT wires are still quite low, but also that they are increasing, suggesting the presence of many extrinsic defects like pores, cracks, incomplete reaction etc. However, the presence of weak-linked current paths in the superconductor is a reasonable inference from the strong degradation in $J_c(H)$ that occurs on application of even weak fields. The main question still to be answered is whether the weak-link behavior is an extrinsic one brought about by the presence of normal phases such as Fe-As that so efficiently wets FBS GBs, or whether it is an intrinsic effect due to the random grain-to-grain, thus generally high-angle misorientation of grains in the wire. Ongoing studies are clearly needed. The prototype Ba-122 coated conductors [106] are very interesting since they utilize the existing YBCO coated conductor template technology and have quickly achieved over 1 MA/cm$^2$. The wide availability of such templates makes experiments in multiple laboratories quite feasible, so we may expect many more experiments in the near future.

## IV Discussion and Summary Reflections

Since we are still just 3 years after the first report of superconductivity above 20 K in an FBS [107] in what has turned out to be a very large class of Fe-based superconductors incorporating both pnictides and chalcogenides, it may be too early to issue a definitive verdict on the intrinsic properties of grain boundaries in the Fe-based superconductors. However, it is clear that the preponderance of evidence is that weak links and low critical current density are associated with all except very low angle grain boundaries. Such behavior poses a major challenge, since there is abundant evidence to show that FBS materials could otherwise be of great potential interest for applications because they have $T_c$ values (25-56K) much higher than any practical Nb-based superconductor (9 K for Nb-Ti and 18 K for Nb$_3$Sn) [7],[8], upper critical fields that are much higher (more than 100 T for some FBS [108] as compared to 15 T for Nb-Ti and 30 T for Nb$_3$Sn), $H_{c2}$ anisotropies that range from a high of about 7 in Sm-1111 to less than 2 in Co-doped Ba-122 (as compared to 1 for Nb-based materials and 5 for YBCO, the least anisotropic cuprate), and intragrain $J_c$ values that exceed $10^6$ A/cm$^2$ at self-field and more than $10^5$ A/cm$^2$ in fields of 20 T. Moreover, it is now becoming clear that Co-doped Ba-122 can accept a very high density of strong correlated pins that yield very high $J_c$ along all axes [109]. Although there may be concerns about the potential toxicity of As, conductors are likely to keep it well contained. There are no particularly expensive elements needed to create FBS. This means that only the poor connectivity of present polycrystalline forms of FBS stands as a major roadblock to their applications. The question that we have tried to address in this review is whether the problem is an intrinsic one that is tied to the very nature of the doping of a non-superconducting compound approach to developing high $T_c$ or whether it is a problem that comes from the great difficulty of mastering the synthesis of complex phases in polycrystalline forms. Understanding the answer to this question is important both for applications of FBS and, in a broader sense, for the applications potential of any superconductor of potentially high $T_c$ that is created by doping carriers into a parent compound that is not superconducting [110,111].

Grain boundaries in the presently used metallic, low-$T_c$ superconductors were addressed in section II.1. The key point that provides context to the discussion of FBS grain boundaries is that Nb-Ti, Nb$_3$Sn, MgB$_2$ and the Chevrel phases comprise a class of important superconductors in which grain boundaries are *beneficial* to the development of high $J_c$. Grain boundaries in these materials allow superconductivity to become a "*little*" depressed, yielding small depressions of the order parameter at the grain boundary that allow vortices to be pinned, hence raising $J_c$, rather than so depressed that the GB becomes a weak link that significantly depresses $J_c$. In these low-$T_c$ materials, the greater the GB density, the higher is the long-range $J_c$. Yet cuprates and now FBS seems to be classes of materials in which grain boundaries disconnect, rather than connect the superconducting grains. Because the principal route to higher $T_c$ is now thought to be the finding of complex compounds with many incipient interactions that can be tuned by small doping additions, it ought to be a major concern of superconductivity research to understand exactly what are the differences between the grain boundaries in high carrier density materials like Nb47wt.%Ti, Nb$_3$Sn, SnMo$_6$S$_8$ or MgB$_2$ and the much lower carrier density compounds like cuprates and FBS. Section II.2 indeed explored the consequences of low carrier density, an unconventional order parameter, and proximity to a parent anti-ferromagnetic state, finding that these are conditions that allow disruption of superconductivity at regions of structural disorder, such as occurs at grain boundaries. One conclusion of this discussion is that FBS may be similar to the cuprates, as far as the depression of the superconducting order parameter at grain boundaries is concerned. The small coherence lengths of FBS and cuprates (actually not much smaller than Nb$_3$Sn) may well be less important than their low carrier densities, large Thomas-Fermi screening lengths and competing superconducting and anti-ferromagnetic orders. Compared to Nb$_3$Sn, which can tolerate and actually benefit from several monolayers of normal Cu segregation at its grain boundaries, even clean FBS and cuprate grain boundaries appear to show depressed superconducting properties.

When superconducting pairing occurs on the nanometer scale, it also becomes hard to define what the local structure and properties are. The particular role played by epitaxially grown thin film bicrystals was addressed in section III.1. Concerning the generality of the conclusions made there, it does need to be said that such films have so far been grown only by 2 groups and only for Co-doped Ba-122. Although there is a slight difference of view between the two studies, actually both studies agree that larger angle [001] tilt bicrystals are weak linked, differing only in what might be the critical angle at which GBs cause the intergranular $J_c$ to fall below the intragrain $J_c$. In one report $\theta_c$ is about 3°, in the other 9°. What is not in doubt is the weak link nature of higher angle [001] tilt grain boundaries with $\theta \gg 10°$. Since most FBS can only be made as bulk polycrystals or single crystals, separation of grain and grain boundary properties remains very important, although difficult. The techniques for such experiments were reviewed in section II.4, following a discussion of single crystal properties in section II.3. Section III.2 discussed the evidence in support of a conclusion that not just Co-doped Ba-122 grain boundaries are weak linked, but that this is also true for carefully made 1111 FBS too. Section III.4 addresses the properties of powder-in-tube wires where so far the evidence for weak link behavior is overwhelming. Wires made of 1111 and 122 show very similar behavior too.

Where then does this strong preponderance of data leave understanding of FBS grain boundaries? It is first important to point out that the difficulties involved in performing the measurements mean that only a limited number of the FBS compounds have had their grain boundary properties characterized. While in the cuprates, the GB behavior is essentially the same across all members, certain cuprates, such as those based on Bi, are easily produced in forms which evade, principally by texture development, rather than overcome the grain boundary problem. The use of textured coated conductor templates for Co-doped $BaFe_2As_2$ shows that this technology can be applied to FBS too. Indeed $J_c$ values that exceed the normally accepted threshold of $10^5$ A/cm$^2$ have been made in this way.

A final point to make for applications is that a multifilamentary wire form conductor is much preferred to a tape for magnet construction. In this respect, we should conclude with one important sign that FBS are indeed not as weak-linked as cuprates, in spite of the lengthy discussion in section II.2 of *how they could be much worse.* In fact the inter-granular $J_c$ of randomly oriented FBS polycrystals is significantly higher than in cuprates, as evidenced by transport $J_c$ values of $10^3$–$10^4$ A/cm$^2$ which have been reported in randomly oriented FBS bulks [69] and wires [103,112], as compared to the typical ~100 A/cm$^2$ of random cuprate bulks. Of course the parent state of FBS is a poor metal, rather than the insulating AF state of cuprates and we also know that overdoping, that is increasing the carrier density, is a good way to enhance $J_{cgb}$ in cuprates. Even if there are many similarities between cuprates and FBS, what may matter for applications, and what is certainly interesting scientifically, are the small differences within the similarities. The fact that a macroscopically untextured, fine-filament (~15 μm) round wire of Bi-2212 is possible [113] because Bi-2212 can be effectively overdoped suggests that small differences do matter. It is our hope that the small differences within various classes of FBS compounds will get the attention that they deserve and allow an important breakout of these materials to applications.

# Acknowledgements

The authors are grateful to many colleagues for discussions, preprints, reprints and access to previously published figures. Special thanks are due to Dmytro Abraimov, Jan Jaroszynski, Jianyi Jiang, Fumitake Kametani, Anatoly Polyanskii, and Jeremy Weiss at the National High Magnetic Field Laboratory, Sanghan Lee at the University of Wisconsin, and Marina Putti and colleagues at the University of Genova, Italy. Work on FBS at the NHMFL was supported in earlier times by the AFOSR, by the NSF/DMR-0084173, by the NSF/DMR-1006584, and by the State of Florida.